\documentclass[aps,prd,a4paper,preprintnumbers,amsmath,amssymb,superscriptaddress,twocolumn,floatfix,showpacs]{revtex4}
\usepackage{amssymb}
\usepackage{float,graphicx}
\usepackage{subfigure}

\newcommand{\Oo}{{\cal O}}
\newcommand{\dd}{{\rm d}}

\begin{document}

\title{Evolution of the Chameleon Scalar Field in the Early Universe}
\author{David F. Mota}
\email{D.F.Mota@astro.uio.no}
\affiliation{Institute of Theoretical Astrophysics, University of Oslo, 0315 Oslo, Norway}
\author{Camilla A.O. Schelpe}
\email{C.A.O.Schelpe@damtp.cam.ac.uk}
\affiliation{Department of Applied Mathematics and Theoretical Physics,
Centre for Mathematical Sciences,  Cambridge CB3 0WA, United Kingdom}
\begin{abstract}
In order to satisfy limits on the allowed variation of particle masses from big bang nucleosynthesis (BBN) until today, the chameleon scalar field is required to reach its attractor solution early on in its cosmological evolution. Brax et al. (2004) have shown this to be possible for certain specific initial conditions on the chameleon field at the end of inflation. However the extreme fine-tuning necessary to achieve this, poses a problem if the chameleon is to be viewed a natural candidate for dark energy. In this article we revisit the behaviour of the chameleon in the early Universe, including the additional coupling to electromagnetism proposed by Brax et al. (2011). Solving the chameleon evolution equations in the presence of a primordial magnetic field, we find that the strict initial conditions on the chameleon field at the end of inflation can be relaxed, and we determine the associated lower bound on the strength of the primordial magnetic field. 

\end{abstract}
\maketitle

\section{Introduction}\label{intro}

Gravity theories which extend General Relativity by introducing a new degree of freedom have received increased attention lately due to combined motivation coming from high-energy physics, cosmology and astrophysics; see \cite{Jain10} for a recent review. If they claim to account for dark energy or dark matter, then they are only valid if they pass the many weak-field limit tests. For many this is only possible if they have a chameleon mechanism \cite{Jain10,Khoury03,Li,Sotiriou,Durrer}. Such mechanisms have the effective mass of the scalar degree of freedom being a function of the curvature (or energy density) of the local environment, so that in effect the mass is large at Solar System and terrestrial curvatures and densities, but small at cosmological curvatures and densities.

The original chameleon model was formulated by Khoury and Weltman \cite{Khoury04,Khoury03} as a scalar field model for dynamical dark energy. It is a scalar-tensor theory in which the new degree of freedom is identified with the chameleon field in the Einstein frame. The interaction of the chameleon $\phi$ to matter is through the conformal metric $A_{\rm m}^2(\phi)g_{\mu\nu}$. It was later shown in \cite{Brax10} that the coupling of the chameleon to charged matter naturally generates a direct coupling between the chameleon and electromagnetic field of the form $A_F(\phi)F_{\mu\nu}F^{\mu\nu}$. This extra term should not be neglected in these models. The interaction between the chameleon and electromagnetic fields leads to conversion between chameleons and photons in the presence of a magnetic field, which has the potential to alter the intensity and polarization of radiation passing through magnetic regions. The astrophysical and laboratory tests for this direct chameleon-photon interaction have been investigated extensively \cite{Brax0703,Brax0707,Burrage07,Burrage08,Li2,Mota,Davis09,Davis10,gies,Schelpe10,Steffen10}. We discuss the chameleon model in more detail in Section \ref{sec:TheChameleonModel}.  

The early Universe behaviour of the chameleon, considering only the interaction of the chameleon to matter species, was analysed by Brax et al. in 2004 \cite{Brax05}. They assumed that the chameleon field was generated at some phase transition during inflation, and that at the end of inflation the field value was left at some arbitrary position within the effective potential. The chameleon then rolls to the minimum of its effective potential and in some cases oscillates about it. Brax et al. \cite{Brax05} assumed a 10\% limit on the variation of particle masses from big bang nucleosynthesis (BBN) until today which requires the maximum amplitude of the chameleon oscillations to be below a certain threshold at BBN. Their analysis, which neglected the electromagnetic interaction, found that the chameleon field initial conditions needed to be strongly fine-tuned if the chameleon's approach to the minimum were to satisfy the constraints placed on the field at BBN. In this article we re-analyse the early Universe behaviour of the chameleon, but this time including the direct coupling between the chameleon and electromagnetic fields. We find that a primordial magnetic field (PMF), produced from some symmetry-breaking phase during inflation, can drive the chameleon towards the minimum of its potential more quickly. Thus satisfying the constraints at BBN for a much wider range of initial conditions.

The existence of a large-scale PMF in the early universe is a matter of current debate. It has so far gone undetected by observations and only upper bounds at the $\mathcal{O}({\rm nG})$ level have been placed on its magnitude. Theories explaining the origin of the PMF are speculative, but it has been suggested that a large-scale magnetic field could be produced during inflation if the conformal invariance of the electromagnetic field is broken \citep{Bertolami98,Dimopoulos02,Gasperini95,Kahniashvili10,Kunze10,Ratra92,Turner87}. See \cite{Kandus11} for a recent review. A PMF is often invoked to explain the order $\mu\mathrm{G}$ magnetic fields observed in nearly all galaxies and galaxy clusters. The origin of these galactic magnetic fields is unclear, but a natural argument is that they develop by some form of amplification from a pre-galactic cosmological magnetic field. Two popular formation scenarios are either some exponential dynamo mechanism which amplifies a very small seed field of order $10^{-30}\mathrm{G}$ as the galaxy evolves, or the adiabatic collapse of a larger existing cosmological field of order $(10^{-10}\text{--}10^{-9})\mathrm{G}$. See \citep{Carilli01,Kronberg94,Kulsrud07,Widrow02,Widrow11} for reviews on the subject. The generic model for the primordial magnetic field is of a stochastic field parameterized by a power-law power spectrum, $P(k)=A_{B}k^{n_{B}}$, up to a cut-off scale $k<k_D$, where $A_B$ is some normalization constant. The spectral index $n_B$ must be greater than $-3$ to prevent infrared divergences in the integral over the power spectrum at long wavelengths. Normalization of the power spectrum is achieved by convolving the magnetic field with a Gaussian smoothing kernel of comoving
radius $\lambda_{B}$ \citep{Kahniashvili08,Kosowsky05}. Constraints on the PMF are then derived in terms of the mean-field amplitude $B_{\lambda}$ of the smoothed field, the smoothing scale $\lambda_{B}$ and the spectral index $n_B$. Kahniashvili et al. \cite {Kahniashvili08} placed bounds on the magnitude of the PMF from comparing the WMAP 5-year data to predicted CMB power spectra that included Faraday rotation effects induced by a PMF. They found that the upper limit on the mean-field amplitude of the magnetic field on a comoving length scale of $\lambda_{B}=1\,\mathrm{Mpc}$ was in the range $6\times10^{-8}$ to $2\times10^{-6}\mathrm{G}$ ($95\%\:\mathrm{CL}$) for a spectral index $n_B=-2.9$ to $-1$. This range for the spectral index was based on the likely formation scenarios for the primordial magnetic field \citep{Bertolami98,Gasperini95,Ratra92,Turner87} and current exclusion bounds on the 
spectral index \citep{Caprini01}. More recent work analysing the WMAP 5-year data in combination with other CMB experiments such as ACBAR, CBI and QUAD have placed tighter constraints on the amplitude of the PMF with $B_{1\,\mathrm{Mpc}}<2.98\times 10^{-9}\mathrm{G}$ and spectral index $n_B<-0.25$ ($95\%\:\mathrm{CL}$) \citep{Yamazaki10}. An analysis of the latest WMAP 7-year data by \cite{Paoletti10} derived upper bounds of $B_{1\,\mathrm{Mpc}}<5.0\times 10^{-9}\mathrm{G}$ and $n_B<-0.12$ ($95\%\:\mathrm{CL}$).

This paper is organized as follows: we begin in Section \ref{sec:TheChameleonModel} by introducing the chameleon scalar field model and deriving the equations governing the chameleon behaviour in the early Universe. Section \ref{sec:Evolution} is dedicated to finding a semi-analytic solution to the evolution of the chameleon for a range of initial field values at the end of inflation. We discuss our results in Section \ref{sec:Results}. We determine the implied bounds on the primordial magnetic field and chameleon parameters from satisfying constraints at BBN in Section \ref{BBNconstraints}, and conclude with a summary of our findings in Section \ref{EarlyUniverseDiscussion}.


\section{The Chameleon Scalar Field Model}\label{sec:TheChameleonModel}

In this section we outline the chameleon scalar field model, and derive the equations governing the chameleon's cosmological evolution. This analysis includes the direct coupling between the chameleon and electromagnetic fields. We consider the effects of a background primordial magnetic field and ignore the second order contribution from the interaction of the chameleon with electromagnetic radiation. 

The action describing the chameleon scalar field model is that of a generalized scalar-tensor theory:
\begin{eqnarray}
\mathcal{S} & = & \int\mathrm{d}^{4}x\,\sqrt{-g}\left(\frac{1}{2}M_{Pl}^{2}\mathcal{R}-\frac{1}{2}g^{\mu\nu}\partial_{\mu}\phi\partial_{\nu}\phi\right.\nonumber \\ 
& & \left. - V(\phi)-\frac{1}{4}A_{F}(\phi)F_{\mu\nu}F^{\mu\nu}\right)\nonumber\\
& & +\int\mathrm{d}^{4}x\, L_{\mathrm{matter}}\left(\psi^{(i)},\, A_{i}^{2}(\phi)g_{\mu\nu}\right)\,,\label{eq:action}
\end{eqnarray}
where $M_{\rm Pl}=1/\sqrt{8\pi G}$ is the reduced Planck mass, and we have explicitly included the kinetic term of electromagnetism outside the matter Lagrangian. The above theory is expressed in the Einstein conformal frame, in which the scalar field $\phi$ is minimally coupled to gravity. In the Jordan frame, described by the metric $g_{\mu\nu}^{(i)}\equiv A_{i}^{2}(\phi)g_{\mu\nu}$, the matter fields $\psi^{(i)}$ are minimally coupled to gravity and are independent of $\phi$; however in the Einstein frame the different matter species couple to the scalar field through the conformal metric $A_{i}^{2}(\phi)g_{\mu\nu}$. It is standard practice to assume the chameleon couples to the different matter species equally so $A_{i}(\phi) = A_{\rm m}(\phi)$.  The coupling between the chameleon and electromagnetic gauge fields is determined by the function $A_{F}(\phi)$. Although the strength of the chameleon interaction to matter and electromagnetism is likely to be of similar magnitude, we do not require it. The coupling functions $A_{\rm m}(\phi)$ and $A_{\rm F}(\phi)$ are generically taken to be of exponential form: $\exp(\phi/M)$ and $\exp(\phi/M_{\rm F})$ respectively. For most situations the chameleon is in the weak field limit, $\phi\ll M,M_{\rm F}$, and so the coupling functions can be Taylor expanded: 
\begin{eqnarray*}
A_{\rm m}  &\approx& 1 + \phi/M \\
A_{F} &\approx& 1 +\phi/M_{\rm F},
\end{eqnarray*} 
where $1/M$ and $1/M_{\rm F}$ describe the strength of the chameleon to matter and photon interactions respectively. 

In this analysis, we take the Universe to be described by a Friedman-Robertson-Walker (FRW) metric and assume approximate spatial flatness, so that
\begin{equation}
\mathrm{d}s^{2} = -\mathrm{d}t^{2}+a^{2}(t)\left(\mathrm{d}x^{2}+\mathrm{d}y^{2}+\mathrm{d}z^{2}\right), \label{eq:metric}
\end{equation}
where $t$ is cosmological time and $a(t)$ is the time-dependent
scale factor describing the expansion, normalized to $a_{0}=1$ today. 

The equation of motion for the $\phi$ field comes from varying $\mathcal{S}$
with respect to $\delta\phi$: 
\[\square\phi=V^{\prime}(\phi)+\frac{1}{4}F^{2}A_{F}^{\prime}(\phi)-\frac{1}{\sqrt{-g}}\frac{\delta L_{\mathrm{matter}}}{\delta\phi},\]
where $\square\phi\equiv\frac{1}{\sqrt{-g}}\partial_{\mu}\left(\sqrt{-g}g^{\mu\nu}\partial_{\nu}\phi\right)$, and for the metric in equation (\ref{eq:metric}), $$\square\phi= -\ddot{\phi}-3H\dot{\phi}+\frac{1}{a^2}\nabla^2\phi,$$ where $H(t)\equiv\dot{a}/a$ is the Hubble expansion rate.
The stress-energy tensor in the Jordan frame is defined by
\begin{equation*}
T_{\mu\nu}^{(i)} \equiv \frac{-2}{\sqrt{-g_{(i)}}}\frac{\delta L_{\mathrm{matter}}}{\delta g_{(i)}^{\mu\nu}},
\end{equation*}
in which particle masses are constant and independent of $\phi$. This leads to \[
\square\phi=V^{\prime}(\phi)+\frac{1}{4}F^{2}A_{F}^{\prime}(\phi)-A_{m}^{3}(\phi)A_{m}^{\prime}(\phi)g_{(i)}^{\mu\nu}T_{\mu\nu}^{(i)}.\]
The stress-energy
tensor for a perfect fluid with density $\rho$ and pressure $p$ is \[
T_{\alpha\beta}=\left(\rho+p\right)u_{\alpha}u_{\beta}+pg_{\alpha\beta},\]
where the 4-velocity of the cosmological fluid is $u_{\alpha}=\left(1,\underline{\boldsymbol{0}}\right)$ in this metric, and so $T^{\mu\nu}=\mathrm{diag}\left(\rho,\, a^{-2}p,\, a^{-2}p,\, a^{-2}p\right)$. The contraction of the stress-energy tensor is then $T_{\mu}^{\mu}=-\rho + 3p$.
The energy densities in the physical Einstein frame, defined by $T_{\mu\nu} \equiv \frac{-2}{\sqrt{-g}}\frac{\delta L_{\mathrm{matter}}}{\delta g^{\mu\nu}}$, are related to those in the Jordan frame by $T_{\mu}^{\mu}=A_{m}^{4}T_{\mu}^{\mu\,(i)}$. Thus
\begin{equation}
\square\phi=V^{\prime}(\phi)+\frac{1}{4}F^{2}A_{F}^{\prime}(\phi)-T_{\mu}^{\mu}A_{m}^{-1}(\phi)A_{m}^{\prime}(\phi). \label{eq:chameleonEqn}
\end{equation}
In general $\phi$ is in the weak-field limit and so we can approximate $A_{m}^{-1}(\phi)\approx 1$. 
This allows us to define an effective potential that the chameleon moves in,
\begin{equation}
V_{\mathrm{eff}}(\phi) \equiv V(\phi)+\frac{1}{4}F^{2}A_{F}(\phi)-T_{\mu}^{\mu}A_{m}(\phi).\label{eq:Veff}
\end{equation}
These equations determine the behaviour of the chameleon in the early Universe in the presence of a primordial magnetic field. 
In general $F^{2}=2\left(\vert B\vert^{2}-\vert E\vert^{2}\right)$ and $T_{\mu}^{\mu}=-\rho_{\rm m}$, where $E$ and $B$ are the background electric and magnetic fields respectively and $\rho_{\rm m}$ is the density of the surrounding matter. 
The mass of small perturbations, $m_{\phi}$, in the chameleon field is given by
 \begin{equation}
m_{\phi}^{2}\equiv V_{,\phi\phi}^{\mathrm{eff}}\left(\phi_{\mathrm{min}};\,\bar{F}^{2},\,\bar{\rho}_{\rm m}\right)\,,\label{eq:chameleon_mass}
\end{equation}
where $\bar{F}^{\mu\nu}$ and $\bar{\rho}_{\rm m}$ are the background values of the fields, and $\phi_{\rm min}$ is the minimum of the effective potential. 

The Friedmann equations governing the expansion of the Universe are also modified by the presence of chameleon dark energy:
\begin{equation*}
3M_{Pl}^{2}\left(\frac{\dot{a}}{a}\right)^{2} = \rho_{\phi}+\rho_{EM}+\rho,
\end{equation*}
\begin{equation*} 
6M_{Pl}^{2}\left(\frac{\ddot{a}}{a} +\left(\frac{\dot{a}}{a}\right)^2 \right)=  \left(\rho_{\phi}-3p_{\phi}\right)+\left(\rho-3p\right),
\end{equation*}
where we have defined the energy density and pressure in the scalar field, neglecting spatial variations in $\phi$, as $\rho_{\phi} \equiv\dot{\phi}^{2}/2+V(\phi)$ and $p_{\phi}\equiv\dot{\phi}^{2}/2-V(\phi)$ respectively, and defined the conserved energy density in the electromagnetic field by $\rho_{EM}\equiv A_{F}(\phi)\left(\vert B\vert^{2}+\vert E\vert^{2}\right)/2$.
Combining these two equations leads to the conservation equation for each species,
\begin{equation}
\rho_{i} = \rho_{i0}a^{-3(1+w_{i})},\label{eq:energy_conservation}
\end{equation}
where $p=w_{i}\rho$ defines the equation of state parameter $w_i$. For radiation we have $w=1/3$, while for non-relativistic
matter $w=0$. 
In general the energy density in the background electromagnetic fields is much less than the energy density of relativistic matter in the Universe, $\rho_{\rm r}$, and so   
\begin{equation}
3M_{Pl}^{2}H^2(a) = \rho_{\phi}+\rho_{\rm m0}a^{-3}+\rho_{\rm r0}a^{-4}, \label{eq: 4}
\end{equation}
which is the standard expression for the Hubble expansion in the presence of dynamical dark energy.  

Equations (\ref{eq:chameleonEqn}) and (\ref{eq: 4}) form the starting point for our analysis of the chameleon evolution in the early Universe. The remainder of this section is dedicated to a brief summary of the properties of the chameleon model, and the experimental bounds that have been placed on its parameters.  

\begin{figure}[htb!]
\begin{centering}
\includegraphics[clip,width=7.5cm]{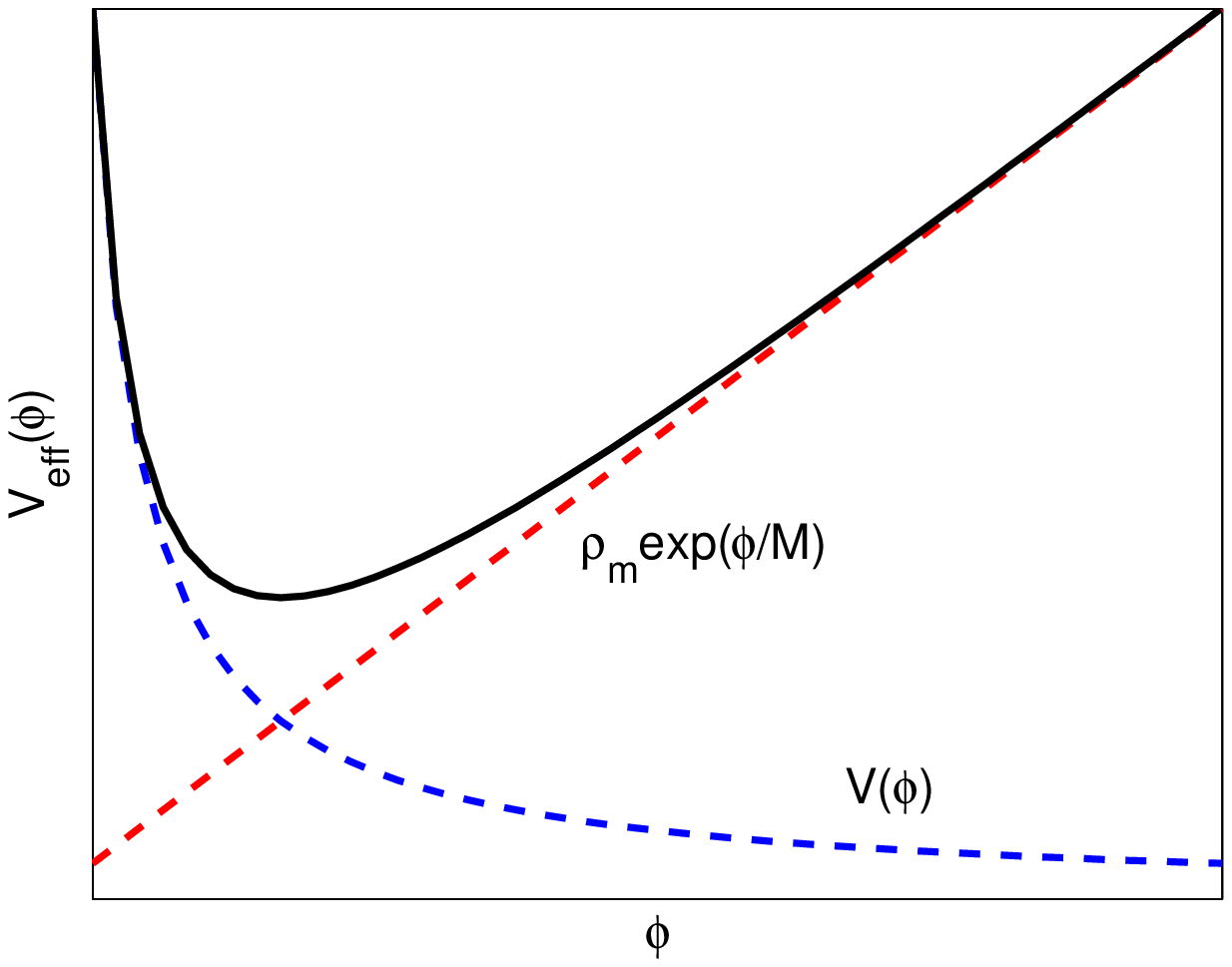}
\includegraphics[clip,width=7.5cm]{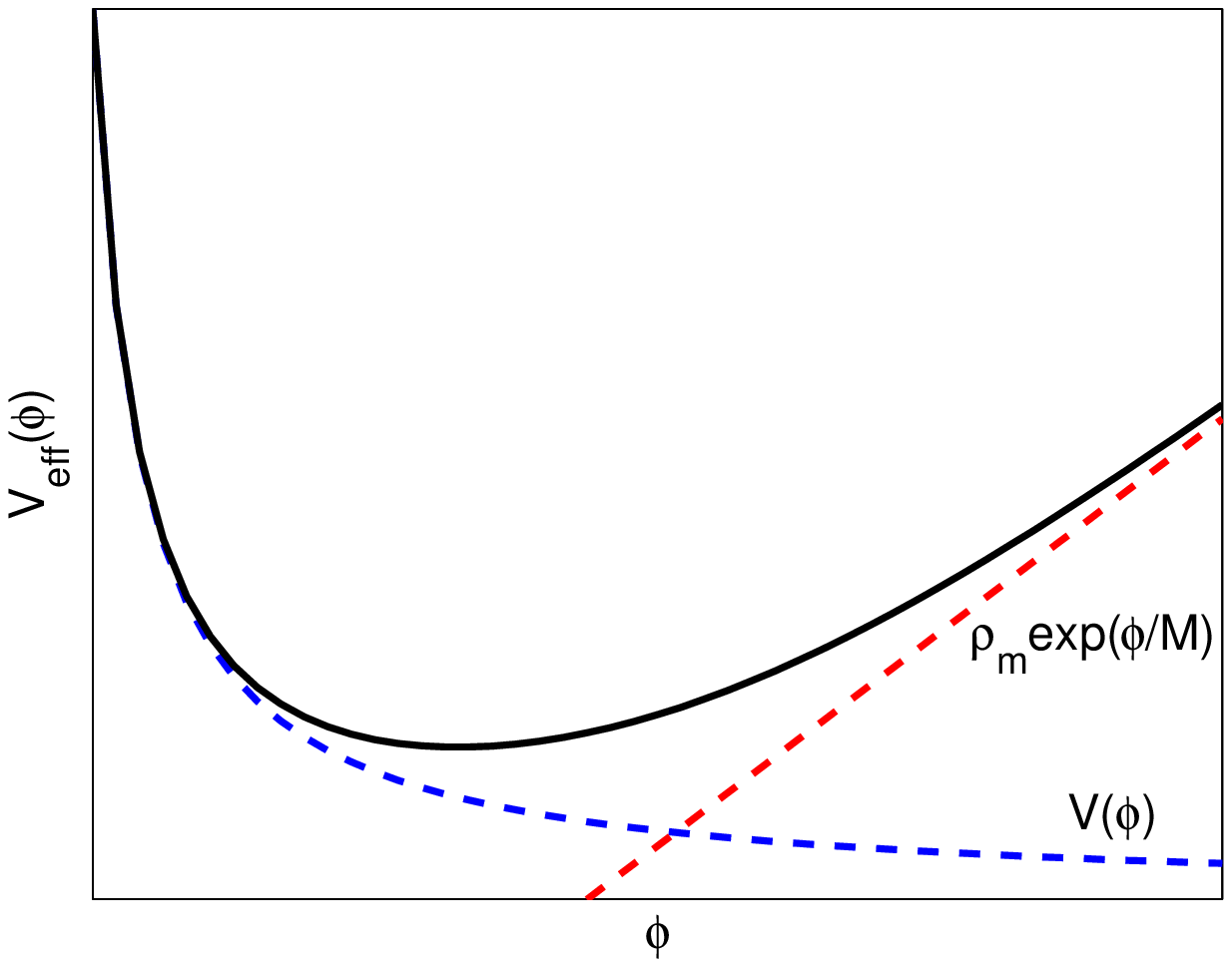}

\par\end{centering}

\caption[The chameleon effective potential.]{\label{fig:chameleon-potential} The chameleon effective potential,
$V_{\mathrm{eff}}$, is the sum of the scalar potential, $V(\phi)$,
and a density-dependent term. In high density environments (upper plot) the mass of the chameleon is much greater than in low density environments (lower plot).} 

\end{figure}

The self-interaction potential of the scalar field, $V(\phi)$, determines whether
a general scalar-tensor theory is chameleon-like or not. For a chameleon we require the mass of small perturbations about the minimum to depend on the surrounding matter density. This imposes that $V(\phi)$ is of runaway form; in other words it is monotonically decreasing and all of $V$, ($V_{,\phi}/V$) and ($V_{,\phi\phi}/V_{,\phi}$) tend to zero as $\phi$ tends to infinity, and to infinity as $\phi$ tends to zero. A typical choice can be described by
\begin{equation}
V(\phi) = \Lambda_{0}^{4}\exp\left(\frac{\Lambda^{n}}{\phi^{n}}\right)\,,\label{eq:self-interaction potential}
\end{equation}
where $n\sim\mathcal{O}(1)$. This is also the desired form for quintessence models of dark energy \citep{Ratra88}. For the chameleon to be a suitable dark energy candidate, we require
$\Lambda_{0}=\left(2.4\pm0.3\right)\times10^{-3}\mathrm{eV}$, and we assume $\Lambda\sim\mathcal{O}\left(\Lambda_{0}\right)$ for reasons of naturalness. The resulting chameleon effective potential is plotted for two different values of $\rho_{\rm m}$ in Figure \ref{fig:chameleon-potential}.
The shape of the effective potential depends on both the background electromagnetic (EM) fields and the density of the surrounding matter. In this plot we neglect the EM contribution and assume $A_{\rm m}(\phi)\simeq \exp(\phi/M)$.  From this we can see that the curvature of the effective potential at the minimum
increases as the density of the surrounding matter, $\rho_{\rm m}$, increases. Thus, in the low density environments of space the chameleon is very light and can drive the cosmic acceleration, while in high density environments, such as in the laboratory, the chameleon is much heavier and evades detection. 

Non-linear self-interactions in the chameleon model also help the scalar field satisfy fifth-force constraints through a property termed the thin-shell effect \citep{Brax07c,Khoury04}. In the thin-shell effect the bulk of the variation in the $\phi$ field, from its equilibrium value inside a test body to its equilibrium value in the surrounding medium, occurs in a narrow band at the surface of the body. The acceleration felt by a test mass from the force mediated by the chameleon particle is proportional to $\frac{1}{M}\boldsymbol{\nabla}\phi$. Thus for a thin-shelled object, it is as if the chameleon field only interacts with the narrow band of matter near the surface of the test body where the variation in $\phi$ occurs, and the resulting fifth-force mediated by the chameleon is strongly suppressed. In general large, dense objects have thin-shells while smaller objects do not. The strength of the chameleon-matter coupling $1/M$ is constrained by existing laboratory experiments \citep{ChamPrimer,Brax10b,Brax07c,Mota06,Mota07}. The best direct bound comes from particle physics experiments, $M\gtrsim 10^4\mathrm{GeV}$ \citep{Brax07c,Mota06,Mota07}.

The strength of the chameleon to photon interaction $1/M_{\rm F}$ is more tightly constrained. Laser-based laboratory searches for these particles, such as PVLAS \citep{Zavattini05,Zavattini07} and GammeV \citep{GammeV}, place an upper bound on the strength of the photon interaction, $M_{\rm F} \gtrsim 10^{6}\,{\rm GeV}$. However, the effects of chameleon-photon mixing will accumulate most during propagation through large-scale astrophysical magnetic fields, and so these scenarios offer the best testing ground for the theory. The constraint coming from considering the limits on the production of starlight polarization in the galactic magnetic field is $M_{\mathrm{F}} \gtrsim 1.1\times 10^9 \mathrm{GeV}$ \citep{Burrage08}, while measurements of the Sunyaev--Zel'dovich effect in galaxy clusters places a lower bound on $M_{\mathrm{F}}$ in the range $M_{\rm F}\gtrsim\left(0.25\text{--}1.14\right)\times 10^9 \mathrm{GeV}$, depending on the model assumed for the cluster magnetic field \citep{Davis09,Davis10}.


\section{The Chameleon Evolution \label{sec:Evolution}}
In this section we proceed to solve the chameleon evolution equations, developed in the previous section, for a range of initial field values at the end of inflation. This is a non-trivial task since the equations governing the chameleon are highly non-linear. While it is possible to approximate the behaviour of the chameleon near the minimum of its potential by a harmonic oscillator, the field at the end of inflation is likely to lie for away from this point and so the assumption of linearity breaks down. A numerical approach to solving the evolution equations exactly is also extremely difficult due to the very different scales governing the two sides of the effective potential. In this analysis we find a semi-analytical solution to the chameleon evolution, using the shape of the potential to our advantage in making a few simplifying assumptions.    

There are a number of steps necessary to build up a framework for solving the chameleon behaviour: in Section \ref{EvolutionAlongTheAttractor} we determine that an attractor solution exists for the chameleon in which the field rolls along the slowly varying minimum of its effective potential. In Section \ref{ApproachingMin} we solve for the initial trajectory of the field towards the minimum and discuss the necessary assumptions that are applied to its effective potential. In Section \ref{subsec:oscillations} we iterate this solution to solve for the subsequent oscillations of the chameleon. An extra contribution to the chameleon evolution comes from particle species dropping out of thermal equilibrium, which we include in our calculations in Section \ref{sub:kicks}.  

\subsection{Evolution along the Attractor}\label{EvolutionAlongTheAttractor}

The effective potential describing the chameleon behaviour was defined in equation (\ref{eq:Veff}). The minimum of the potential will evolve over time as the energy density in the matter and electromagnetic fields decreases with the expansion of the Universe. We begin our analysis by checking that the minimum provides a viable attractor solution for the chameleon when the electromagnetic interaction is included in the chameleon model. This requires the evolution of the potential to be slow enough for the chameleon to respond to these changes and follow the minimum. We follow a similar procedure to that outlined in \cite{Brax05}.

The shift in the minimum of the potential is as a result of the expansion of the Universe, and so the characteristic time scale for the change is approximately a Hubble time $H^{-1}$. The characteristic response time of the chameleon is the period of oscillations about the minimum which is given by $m_{\phi}^{-1}$. For the chameleon to adjust quickly enough to changes in the position of the minimum, the ratio $m_{\phi}/H$ needs to be greater than one.

The minimum of the effective potential occurs when $V_{\mathrm{eff}}^{\prime}(\phi)=0$, for which
\begin{eqnarray*}
V^{\prime}(\phi_{\mathrm{\min}}) + \frac{\rho_{\mathrm{PMF}}}{M_F} + \frac{\rho_{\mathrm{m}}}{M} = 0,
\end{eqnarray*} 
where we have defined the energy density in the PMF, $\rho_{\rm PMF}\equiv \vert\mathbf{B}\vert^2/2$. We assume $\phi_{\mathrm{min}}\ll M,\, M_{\rm F}$ such that $A_{\rm m}^{\prime}\approx 1/M$ and $A_{\rm F}^{\prime}\approx 1/M_{\rm F}$. For the self-interaction potential given in equation (\ref{eq:self-interaction potential}) we have, 
\begin{equation}
n\Lambda^n \phi_{\mathrm{\min}}^{-(n+1)}V(\phi_{\mathrm{\min}}) = \frac{\rho_{\mathrm{PMF}}}{M_F} + \frac{\rho_{\mathrm{m}}}{M}.\label{eq:phimin}
\end{equation}
Substituting into the expression for the chameleon mass (\ref{eq:chameleon_mass}) and rearranging,
\begin{eqnarray*}
m_{\phi}^2 & = & \frac{1}{\Lambda}\left(\frac{\rho_{\mathrm{PMF}}}{M_F} + \frac{\rho_{\mathrm{m}}}{M}\right)\left((1+n)\frac{\Lambda}{\phi_{\rm min}}+n\left(\frac{\Lambda}{\phi_{\rm min}}\right)^{n+1}\right)\\
&& +\frac{\rho_{\mathrm{PMF}}}{M_F^2} + \frac{\rho_{\mathrm{m}}}{M^2}.
\end{eqnarray*}
Under the assumption $M\sim M_{\rm F}$ and taking $\phi_{\mathrm{min}}\ll M$, we approximate  
\begin{eqnarray*}
\frac{m_{\phi}^2}{H^2} & \approx & \frac{3M_{\rm Pl}^2\left(\Omega_{\rm PMF}+ \Omega_{\rm m}\right)}{\Lambda M} \left((1+n)\frac{\Lambda}{\phi_{\mathrm{min}}} +n\left(\frac{\Lambda}{\phi_{\mathrm{min}}}\right)^{n+1}\right),
\end{eqnarray*}
where $\Omega_{\rm PMF} \equiv \rho_{\rm PMF}/3H^2M_{\rm Pl}^2$ and $\Omega_{\rm m} \equiv \rho_{\rm m}/3H^2M_{\rm Pl}^2$ are the fractional energy densities in the magnetic field and in non-relativistic (matter) degrees of freedom respectively.   
In the limit $\phi_{\mathrm{min}}\ll \Lambda$, 
\begin{eqnarray*}
\frac{m_{\phi}^2}{H^2} &\approx& \frac{3M_{\rm Pl}^2\left(\Omega_{\rm PMF}+ \Omega_{\rm m}\right)}{\Lambda M}n\left(\frac{\Lambda}{\phi_{\mathrm{min}}}\right)^{n+1} \\
& \gtrsim & \frac{3nM_{\rm Pl}^2\left(\Omega_{\mathrm{PMF}}+ \Omega_{\mathrm{m}}\right)}{\Lambda M}.
\end{eqnarray*}
The smallest value of the function, $\Omega_{\rm PMF} + \Omega_{\rm m}$, occurs at early times ($\Omega_{\rm m}\sim 10^{-7}$ at $a\sim10^{-23}$) and for the limiting case of $\Omega_{\rm PMF}=0$. The largest natural value for the mass parameter of the chameleon coupling $M$ is $\mathcal{O}(M_{\rm Pl})$. Taking $\Lambda\sim10^{-3}\text{eV}$ and $n\sim\mathcal{O}(1)$, we expect 
\begin{eqnarray*}
\frac{m_{\phi}^2}{H^2} & \gtrsim & 10^{24}.
\end{eqnarray*}
Similarly for the limit when $\phi_{\mathrm{min}}\gg \Lambda$, 
\begin{eqnarray*}
\frac{m_{\phi}^2}{H^2} & \approx & (1+n)\frac{3M_{\rm Pl}^2\left(\Omega_{\mathrm{PMF}}+ \Omega_{\mathrm{m}}\right)}{\Lambda M}\frac{\Lambda}{\phi_{\mathrm{min}}},
\end{eqnarray*}
and we can approximate $V(\phi_{\mathrm{min}})\approx \Lambda^4$ in equation (\ref{eq:phimin}) to determine $\phi_{\rm min}$. Then, 
\begin{eqnarray*}
\frac{m_{\phi}^2}{H^2} & \approx & \frac{3(1+n)M_{\rm Pl}^2}{\Lambda M}\left(\frac{3M_{\rm Pl}^{2}H^2}{nM\Lambda^{3}}\right)^{\frac{1}{n+1}} \left(\Omega_{\mathrm{PMF}}+ \Omega_{\mathrm{m}}\right)^{\frac{n+2}{n+1}}.
\end{eqnarray*}
The function $\left(\frac{H}{H_0}\right)^2\left(\Omega_{\mathrm{PMF}}+ \Omega_{\mathrm{m}}\right)^{n+2}$ is minimised at late times. It is greater than $10^{-5}$ for all $n\sim\mathcal{O}(1)$. Again taking $M\sim M_{\rm Pl}$ and $\Lambda\sim10^{-3}\text{eV}$, we expect 
\begin{eqnarray*}
\frac{m_{\phi}^2}{H^2} & \gtrsim & 10^{31-\frac{35}{n+1}}.
\end{eqnarray*}
Thus, we can confirm that at all times the ratio $m_{\phi}/H$ is much greater than unity, and so the minimum does provide a viable attractor solution for the chameleon. Once the chameleon has settled to the minimum of its potential it will always be able to track the minimum as the Universe expands.

\subsection{Approaching the Minimum of the Potential}\label{ApproachingMin}

We now turn our attention to the approach of the chameleon to the attractor solution. Although the origin of the chameleon field is speculative, it is reasonable to assume that it is produced at some phase transition during inflation, and that at the end of inflation the field value is left at some arbitrary position within the effective potential. We expect the field to roll to the minimum of the potential, oscillate about the minimum before then settling to its equilibrium value. The presence of a large scale magnetic field provides an additional source term driving the chameleon to its minimum. The Universe becomes a good conductor during the reheating phase at the end of inflation, after which any electric fields in the Universe will be dissipated and the primordial magnetic field becomes frozen-in, so that from then on $\vert\mathbf{B}\vert^{2}\propto a^{-4}$. 

The equations governing the chameleon evolution were derived in Section \ref{sec:TheChameleonModel}. Assuming the field is spatially homogeneous, equation (\ref{eq:chameleonEqn}) simplifies to,
\begin{equation}
-\frac{\mathrm{d}^2\phi}{\mathrm{d}t^2} - 3H\frac{\mathrm{d}\phi}{\mathrm{d}t} = V^{\prime}(\phi) + \rho_{\rm PMF}A_{\mathrm{F}}^{\prime}(\phi) - T_{\mu}^{\mu}A_{\rm m}^{\prime}(\phi), \label{eq:chameleon-t}
\end{equation}
which describes a weakly damped oscillator, where the friction term proportional to $H$ is due to the expansion of the Universe. The evolution of $H(t)$ was given in equation (\ref{eq: 4}). Prior to BBN the Universe is radiation dominated and so we can approximate
\begin{equation}
H^2(a)\simeq H_0\Omega_{\rm r0}a^{-4}, \label{eq:Hrad}
\end{equation}
where $\Omega_{\rm r0}\equiv 8\pi G\rho_{\rm r0}/3H_0^2$ is the fractional energy density in radiation. In general the stress-energy tensor $T_{\mu}^{\mu}=-\rho_{\rm m}$, where $\rho_{\rm m}$ is the energy density in non-relativistic matter. Soon after reheating, all particle species are relativistic and so the driving term in the chameleon effective potential has $T_{\mu}^{\mu}\approx 0$. However, as the Universe cools the particles drop out of thermal equilibrium. This causes a short-lived contribution to $T_{\mu}^{\mu}$ which can significantly affect the chameleon as it rolls to the minimum. We follow \cite{Brax05} and refer to these contributions as `kicks'. The first kick occurs at a redshift of approximately
$z\sim10^{14}$ as the top quark drops out of thermal equilibrium. To start with we neglect this contribution and solve for the initial behaviour of the chameleon. We include the contribution from these kicks in Section \ref{sub:kicks}. 

Neglecting the contribution from the kicks for now, the effective potential characterising the chameleon motion in equation (\ref{eq:Veff}), can be expressed as,  
\begin{equation*}
V_{\rm eff}(\phi) = \Lambda^4 \exp{\left(\frac{\phi}{\Lambda}\right)^{-n}} + \rho_{\mathrm{PMF}}\exp\left(\frac{\phi}{M_{\rm F}}\right) + \rho_{\mathrm{m}}\exp\left(\frac{\phi}{M}\right).
\end{equation*}
Notice that $M,M_{\rm F}\gg\Lambda$ and so the lefthand side of the potential with $\phi < \phi_{\rm min}$ is many orders of magnitude steeper than that for $\phi > \phi_{\rm min}$. The exponential decay of the functions means that for $\phi < \phi_{\rm min}$ the potential is almost entirely dominated by $V(\phi)$, while for $\phi > \phi_{\rm min}$ the dominant driving force is from the $\rho_{\rm m}$ and $\rho_{\rm PMF}$ terms. This strong asymmetry in the chameleon effective potential allows us to make a few simplifying assumptions so that the chameleon evolution can be determined analytically. 

Whenever $\phi > \phi_{\rm min}$ we assume $V(\phi)$ is negligible. Approximating $A_{\rm m}(\phi),\,A_{\rm F}(\phi)\approx1$, we can solve for the evolution along this side of the potential so that equation (\ref{eq:chameleon-t}) becomes
\begin{equation*}
\frac{1}{a^{2}}H\frac{\dd}{\dd a}\left(a^{4}H\frac{\dd\phi}{\dd a}\right) \simeq -\frac{\rho_{\rm PMF0}}{M_{\mathrm{F}}}a^{-4} -\frac{\rho_{\rm m0}}{M}a^{-3},
\end{equation*}
where we have converted from cosmological time to the scale factor using the relation $H(a)\equiv\dot{a}/a$. The contribution from (non-relativistic) matter, $\rho_{\rm m}$, will be negligible prior to BBN and can be omitted. Substituting equation (\ref{eq:Hrad}) for the evolution of the Hubble expansion, we find
\begin{eqnarray}
\frac{{\rm d}\phi}{{\rm d}a} & = & -\beta a^{-1}+Aa^{-2},\nonumber\\
\phi(a) & = & -\beta\log a -Aa^{-1}+B, \label{eq:EMside}
\end{eqnarray}
where 
\begin{equation}
\beta\equiv \rho_{\rm PMF0}/\Omega_{\rm r0}H_0^2M_{\mathrm{F}}. \label{eq:beta}
\end{equation}
The constants $A$ and $B$ are determined by the initial conditions. 

Conversely, when $\phi<\phi_{\rm min}$ the $\rho_{\rm PMF}$ term is negligible compared to $V(\phi)$. The friction term proportional to $H$ in equation (\ref{eq:chameleon-t}) is also negligible since the steepness of the potential dominates. The roll up or down on this side of the potential will be very rapid and we treat it as instantaneous. Thus as the chameleon approaches from larger field values, $V(\phi)$ acts as a perfect elastic collision. 

We assume the chameleon starts from rest at some initial value $\phi_{i}$ at the end of inflation, $a_{i}\sim10^{-23}$. For starting values with $\phi_{i}\lesssim\phi_{\rm min}$, the chameleon falls very quickly to the minimum and overshoots to the other side of the potential before coming to a halt. The self-interaction potential $V(\phi)$ dominates the initial roll to the minimum. From equation (\ref{eq:chameleon-t}), 
\begin{eqnarray*}
\frac{{\rm d}^{2}\phi}{{\rm d}t^{2}} & \simeq & -V^{\prime}(\phi)
\end{eqnarray*}
and so the velocity as it shoots past the minimum is
$\phi_{,\, t}\simeq \sqrt{2V(\phi_{i})}$ (since $V(\phi_i)\gg V(\phi_{\rm min})$). We treat
this as instantaneous and occurring at $a=a_{i}$. Beyond the minimum
the evolution is determined by equation (\ref{eq:EMside}). The initial conditions, $\phi(a_i)=\phi_{\mathrm{min}}(a_{i})$
and $\phi_{,\,a}(a_i)=\sqrt{2V(\phi_{i})}/a_{i}H(a_{i})$, determine the constants $A$ and $B$. Substituting, we find that the field comes to a halt at
\begin{eqnarray*}
a_{max} & \simeq & a_{i}+\frac{\sqrt{2V(\phi_{i})}}{\beta H(a_{i})}a_{i},\\
\phi_{max} & = & -\beta\ln\frac{a_{max}}{a_{i}}+\frac{\sqrt{2V(\phi_{i})}}{H(a_{i})},
\end{eqnarray*}
where we have assumed $\phi_{max}\gg\phi_{min}$.
The chameleon will then start to fall back towards the minimum and evolve identically to the case when $\phi_{i}\gg\phi_{\rm min}$,
but with $a_{max}$ and $\phi_{max}$ as the initial conditions. Depending on the size of the background magnetic field and the initial starting value for the chameleon field, either the field falls to the minimum and starts oscillating about it, or the friction term dominates and the field remains frozen at its initial value until the first kick occurs as the top quark drops out of thermal equilibrium.

\subsection{Oscillations about the Minimum}\label{subsec:oscillations}

In general, the velocity of the chameleon when it first reaches the minimum will be very large and cause it to overshoot. It will then oscillate about the minimum with a large amplitude which gradually decays due to the damping term from the Hubble expansion. We apply the approach outlined in the previous section to iterate over multiple oscillations and determine the evolution of the amplitude of the chameleon oscillations. 

Let us consider a single oscillation which starts from rest at $\phi=\phi_1$ at some time $a_1$ with $\phi_1\gg\phi_{\rm min}$. The field is governed by equation (\ref{eq:EMside}) as it rolls to the minimum. Substituting the initial conditions to determine $A$ and $B$ gives
\begin{eqnarray*}
\frac{{\rm d}\phi}{{\rm d}a} & \simeq & -\beta a^{-1}\left(1-\frac{a_{1}}{a}\right),\\
\phi-\phi_{1} & \simeq & -\beta\left(\log\left(\frac{a}{a_{1}}\right)+\frac{a_{1}}{a}-1\right).
\end{eqnarray*}
At the minimum,
\begin{eqnarray}
\left.\frac{{\rm d}\phi}{{\rm d}a}\right|_{\phi_{\rm min}} & \simeq & -\beta a_{\rm min}^{-1}\left(1-\frac{a_{1}}{a_{\rm min}}\right),\nonumber \\
\phi_{\rm min}-\phi_{1} & \simeq & -\beta\left(\log\left(\frac{a_{\rm min}}{a_{1}}\right)+\frac{a_{1}}{a_{\rm min}}-1\right), \label{eq:amin}
\end{eqnarray}
which form the initial conditions for rolling back up. Equation (\ref{eq:amin}) can be inverted to determine $a_{\rm min}$ under the assumption $\phi_{\rm min}\ll \phi_1$. Following the approximation discussed in the previous section, we assume the field undergoes an instantaneous perfect elastic collision at the minimum and rebounds with the same velocity. Substituting into equation (\ref{eq:EMside}), the chameleon as it rolls back to larger field values is governed by
\begin{eqnarray*}
\frac{{\rm d}\phi}{{\rm d}a} & \simeq & \beta a^{-2}\left(2a_{\rm min}- a - a_1\right),\\
\phi-\phi_{1} & \simeq & - \beta \left(\log\frac{a}{a_1}+2\left(\frac{a_1}{a_{\rm min}} + \frac{a_{\rm min}}{a}\right) -3 - \frac{a_{1}}{a}\right),
\end{eqnarray*}
It comes to a halt at the retracement point $\phi_{2}$ at some time $a_2$: 
\begin{eqnarray}
a_{2} & \simeq & 2a_{\rm min}-a_{1},\label{eq:a2}\\
\phi_{1}-\phi_{2} & \simeq & \beta\left[\log\left(\frac{a_{2}}{a_{1}}\right)+2\left(\frac{a_{1}}{a_{\rm min}}-1\right)\right].\label{eq:phi2}
\end{eqnarray}
This evolution from one retracement point $\left(\phi_1,a_1\right)$ to another $\left(\phi_2,a_2\right)$ can be iterated to determine the chameleon behaviour. 

An analytic approximation for the evolution of the retracement point $\phi_{\rm max}$ exists in the limit of fast oscillations when $a_{\mathrm{min}}/a_{1}\approx1+\delta$, $\delta\ll1$. In this limit, equation (\ref{eq:amin}) becomes
\begin{equation*}
\frac{\phi_{1}}{\beta}+1 \simeq 1+\frac{1}{2}\delta^{2}+\mathcal{O}(\delta^{3}),
\end{equation*}
which implies $\delta \simeq \sqrt{2\phi_{1}/\beta}$.
Similarly equations (\ref{eq:a2}) and (\ref{eq:phi2}) can be approximated in the limit of small $\delta$, such that
\begin{eqnarray*}
\Delta a\equiv a_{2}-a_{1} & \simeq & 2\delta a_{1},\\
\Delta\phi_{\max}\equiv \phi_{2}-\phi_{1} & \simeq & -\frac{2}{3}\delta^{3}\beta.
\end{eqnarray*}
In the limit of fast oscillations we can approximate, 
\begin{eqnarray}
\frac{\mathrm{d}\phi_{\mathrm{max}}}{\mathrm{d}a} & \approx & \frac{\Delta\phi_{\mathrm{max}}}{\Delta a}\approx-\frac{2}{3}\frac{\phi_{1}}{a}\nonumber\\
\implies\log\phi_{\mathrm{max}} & \simeq & -\frac{2}{3}\log a+\mathrm{const.}\label{eq:fastOSC}\end{eqnarray}
This power law behaviour is independent of the strength of the magnetic field.
In this regime, the magnetic field strength will dictate how rapidly the chameleon oscillates
but not how long it takes to settle to the minimum.

\subsection{Kicks}\label{sub:kicks}

\begin{figure}
\includegraphics[width=7.5cm]{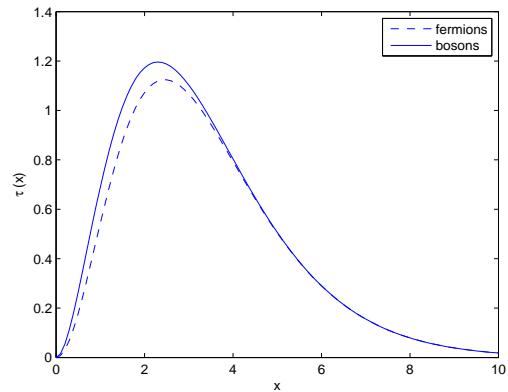}
\caption[The function $\tau(x)$.]{\label{fig:tau} The function $\tau(x)=x^{2}\intop_{x}^{\infty}\mathrm{d}u\frac{\sqrt{u^{2}-x^{2}}}{e^{u}\pm1}$,
where the $\pm$ is for fermions and bosons respectively.}
\end{figure}

In addition to the oscillations described in Section \ref{subsec:oscillations}, the chameleon behaviour is modified by an extra contribution to $T_{\mu}^{\mu}$ as particle species drop out of thermal equilibrium. This manifests itself as a short-lived boost to the driving term on the righthand side of equation (\ref{eq:chameleon-t}). The contribution as each particle, labelled by $k$, goes non-relativistic is
\begin{equation}
T_{\mu}^{\mu\,(k)}= -\frac{45}{\pi^{4}}H^{2}(a)M_{\mathrm{Pl}}^{2}\frac{g_{k}}{g_{\star}(T)}\tau\left(\frac{m_{k}}{T}\right), \label{eq:Tmumu}
\end{equation}
where 
\[
\tau(x)\equiv x^{2}\intop_{x}^{\infty}\mathrm{d}u\frac{\sqrt{u^{2}-x^{2}}}{e^{u}\pm1},
\]
and the $\pm$ sign is for fermions and bosons respectively \citep{Brax05,Damour93,Damour94}. The $\tau$ function is plotted in Figure \ref{fig:tau}. It is approximately $\mathcal{O}(1)$ for $x \sim 1$ and negligible otherwise. 
The temperature of the Universe redshifts as $T=T_0 a^{-1}$. The maximum contribution to $T_{\mu}^{\mu}$ occurs when the temperature of the Universe has cooled sufficiently for it to match the mass of the particle, $m_k \sim T$. The mass of the different species, along with the ratio of the number of degrees of freedom to the effective number of relativistic degrees of freedom,
$g_{k}/g_{\star}$, are listed in Table \ref{tab:degreesoffreedom}. 
\begin{table}
\begin{tabular}{|c|c|c|c|}
\hline 
Particle & mass, $m_k$ & $g_k/g_{\star}$ (at $T=m_k$) & type\tabularnewline
\hline 
$t$ & 173 GeV & 12/106.75 & fermion\tabularnewline
\hline 
$Z$ & 91 GeV & 3/95.25 & boson\tabularnewline
\hline 
$W^{\pm}$ & 80 GeV & 6/92.25 & boson\tabularnewline
\hline 
$b$ & 4 GeV & 12/86.25 & fermion\tabularnewline
\hline 
$\tau$ & 1.7 GeV & 4/75.75 & fermion\tabularnewline
\hline 
$c$ & 1.27 GeV & 12/72.25 & fermion\tabularnewline
\hline 
$\pi$ & 0.14 GeV & 3/17.25 & boson\tabularnewline
\hline 
$\mu$ & 0.105 GeV & 4/14.25 & fermion\tabularnewline
\hline 
$e$ & 0.5 MeV & 4/10.75 & fermion\tabularnewline
\hline 
\end{tabular} 

\caption{\label{tab:degreesoffreedom}List of particle species that provide
a `kick' to the chameleon evolution as they drop out of thermal equilibrium.}

\end{table}

An exact solution to the chameleon evolution with the kicks is not possible. Instead we consider the two limiting cases: when the duration of the kick is very much less than the oscillation period of the chameleon, we can approximate the contribution to $T_{\mu}^{\mu}$ as a delta function similar to the procedure outlined in \cite{Brax05}; on the other hand when the chameleon is oscillating rapidly about the minimum, we can make an adiabatic approximation and assume the function $\tau$ is constant over one oscillation. What is a reasonable estimate for the duration of the kick? Referring to Figure
\ref{fig:tau}, the width of the $\tau$ function runs from approximately
$a/a_{k}=0.1\text{--}10$, beyond which it has dropped to less than 10\%
of its maximum value. We therefore assume the transition between the two regimes occurs when the period of oscillation satisfies $a_2/a_1 \sim 100$,
where $a_2 - a_1 = \Delta a_{\rm osc}$.   

\subsection{Case 1: $\Delta a_{\rm kick} \ll \Delta a_{\rm osc}$}

For the situation when the duration of the kick is much less than the chameleon period of oscillation, we take the contribution to $T_{\mu}^{\mu}$ to be approximated by a delta function. The chameleon spends most of its time during oscillations on the righthand side
of the potential ($\phi>\phi_{\mathrm{min}}$), and so we make the reasonable assumption that the chameleon will be rolling along
this side of the potential when the kick occurs. As before, equation (\ref{eq:chameleon-t}) is approximated by 
\begin{eqnarray*}
\frac{1}{a^{2}}H\frac{\rm d}{{\rm d}a}\left(a^{4}H\frac{{\rm d}\phi}{{\rm d}a}\right) & \simeq & -\frac{\rho_{\rm PMF0}}{M_{\mathrm{F}}}a^{-4} + \frac{1}{M}T_{\mu}^{\mu},
\end{eqnarray*}
and $H(a)$ is given by equation (\ref{eq:Hrad}). Approximating $\tau(x)$ as a delta-function centered on $a=a_{k}$ for each kick, equation (\ref{eq:Tmumu}) becomes
\[
T_{\mu}^{\mu\,(k)}\approx-\frac{45}{\pi^{4}}H^{2}(a)M_{\mathrm{Pl}}^{2}\frac{g_{k}}{g_{\star}(a_k)}a_{k}\Gamma\delta\left(a-a_{k}\right),
\]
where $a_{k}\equiv T_{0}/m_{k}$, and $\Gamma\approx4.9\text{--}4.6$
is the total area under the $\tau(x)$ curve. Note that the delta function satisfies $\delta(x/x_{0}-1)=x_{0}\delta(x-x_{0})$.
Then, defining $\kappa_k\equiv 45 M_{\mathrm{Pl}}(g_{k}/g_{\star})/\pi^{4}M$,
\begin{eqnarray*}
\frac{\rm d}{{\rm d}a}\left(a^{2}\frac{{\rm d}\phi}{{\rm d}a}\right) & \simeq & -\beta -\kappa_k M_{\mathrm{Pl}}\Gamma a_{k}\delta\left(a-a_{k}\right),
\end{eqnarray*}
where as before $\beta\equiv \rho_{\rm PMF0}/\Omega_{\rm r0}H_0^2M_{\mathrm{F}}$. Integrating, we find
\begin{eqnarray*}
\frac{{\rm d}\phi}{{\rm d}a} & \simeq & \begin{cases}
-\beta a^{-1}+Aa^{-2}, & a<a_{k},\\
-\beta a^{-1}+Aa^{-2}-\kappa_k M_{\mathrm{Pl}}\Gamma a_{k}a^{-2}, & a>a_{k},\end{cases}
\end{eqnarray*}
and
\begin{eqnarray*}
\phi(a) & \simeq & \begin{cases}
-\beta\log a-\frac{A}{a}+B, & a<a_{k},\\
-\beta\log a-\frac{A}{a}+B-\kappa_k M_{\mathrm{Pl}}\Gamma\left(1 -\frac{a_k}{a}\right), & a\geq a_{k},\end{cases}
\end{eqnarray*}
where $A$ and $B$ are constants that are determined from the chameleon
evolution prior to the kick. The change in the chameleon from its
free evolution without the kick, is
\begin{equation}
\Delta\phi = -\kappa_k M_{\mathrm{Pl}}\Gamma\left[1-\frac{a_{k}}{a}\right], \label{eq:jumpInPhi}
\end{equation}
which tends to $-\kappa_k M_{\mathrm{Pl}}\Gamma$ in the limit $a\gg a_{k}$. 
The above result assumes the particle drops out
of thermal equilibrium instantaneously relative to the background
evolution, which is only valid when the duration of the kick is much smaller than the oscillation period.

\subsection{Case 2: $\Delta a_{\rm kick} \gg \Delta a_{\rm osc}$}

For the situation when the duration of the kick is significantly greater than the oscillation period of the chameleon, we use an adiabatic approximation treating $\tau(a/a_{k})$ as constant
over one oscillation. We follow the same analysis as in Sections \ref{ApproachingMin} and \ref{subsec:oscillations} but with the extra source term from $T_{\mu}^{\mu}$:
\begin{equation*}
\frac{1}{a^{2}}H\frac{\rm d}{{\rm d}a}\left(a^{4}H\frac{{\rm d}\phi}{{\rm d}a}\right) \simeq - \frac{\rho_{\rm PMF0}}{M_{F}}a^{-4}- H^{2}M_{\mathrm{Pl}}\sum_{k}\kappa_{k}\tau\left(\frac{a_1}{a_k}\right),
\end{equation*}
where $a_1$ is the scale factor at the start of the oscillation and we are assuming $\tau$ does not change significantly from $a_1$ to $a_2$ (the scale factor at the end of the oscillation). Substituting in equation (\ref{eq:Hrad}) for the Hubble expansion, and solving, we find  
\begin{eqnarray}
\frac{{\rm d}\phi}{{\rm d}a} & \simeq & -\beta_{\rm eff} a^{-1}+Aa^{-2},\nonumber\\
\phi(a) & \simeq & -\beta_{\rm eff}\log a -Aa^{-1}+B,\label{eq:adiabaticKicks}
\end{eqnarray}
which is identical to equation (\ref{eq:EMside}) but with $\beta$ replaced by
\[
\beta_{\mathrm{eff}}\equiv\beta+M_{\mathrm{Pl}}\sum_{k}\kappa_{k}\tau\left(\frac{a_1}{a_k}\right).\]
In this regime the effect of the kicks is similar to a temporary increase in the magnetic field strength, and generally boosts the oscillations into
the `fast oscillations' regime described by equation (\ref{eq:fastOSC}).


\section{Results \label{sec:Results}}

\begin{figure*}[t!]
\begin{centering}
\subfigure[$\xi_{B}=1$, $\phi_i=10^5M$, $M=10^{18}{\rm GeV}$.]{\label{fig:a}\includegraphics[bb=3.5cm 9.1cm 17cm 19.5cm,clip,width=7cm]{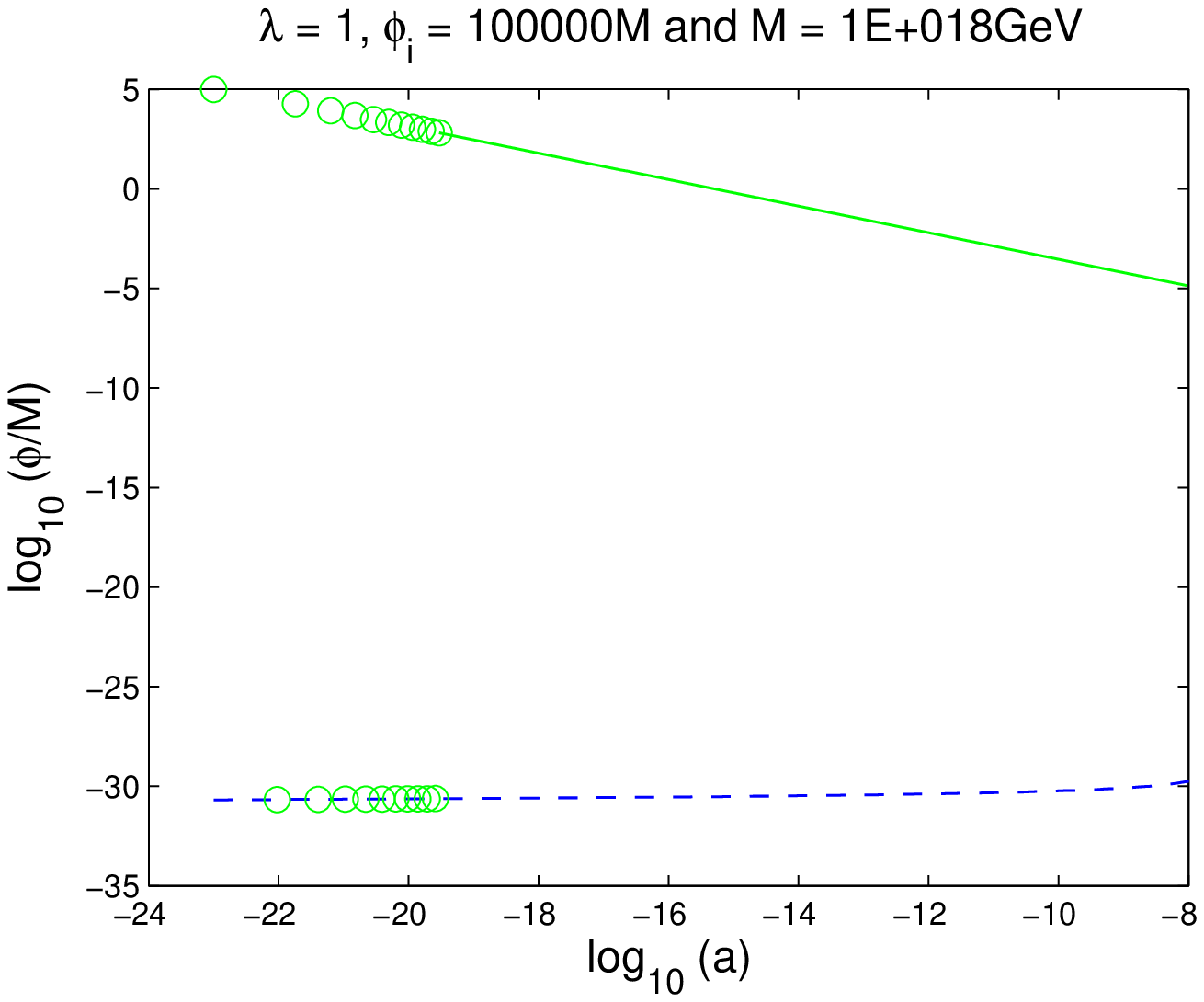}}
\subfigure[$\xi_{B}=1$, $\phi_i=10^6M$, $M=10^{18}{\rm GeV}$.]{\label{fig:b}\includegraphics[bb=3.5cm 9.1cm 17cm 19.5cm,clip,width=7cm]{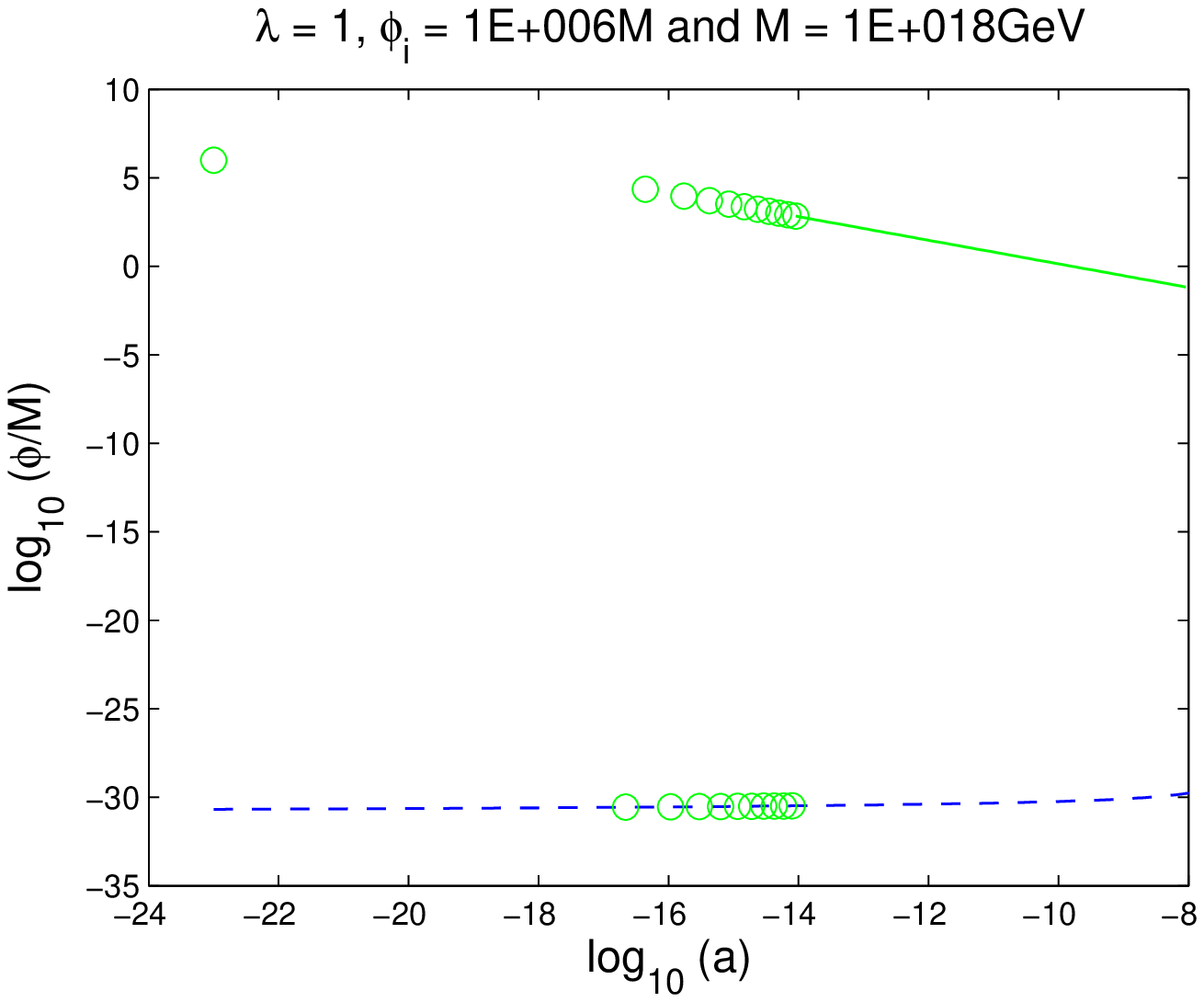}}
\subfigure[$\xi_{B}=1$, $\phi_i=2.5\times10^6M$, $M=10^{18}{\rm GeV}$.]{\label{fig:c}\includegraphics[bb=3.2cm 9.1cm 17.2cm 19.5cm,clip,width=7cm]{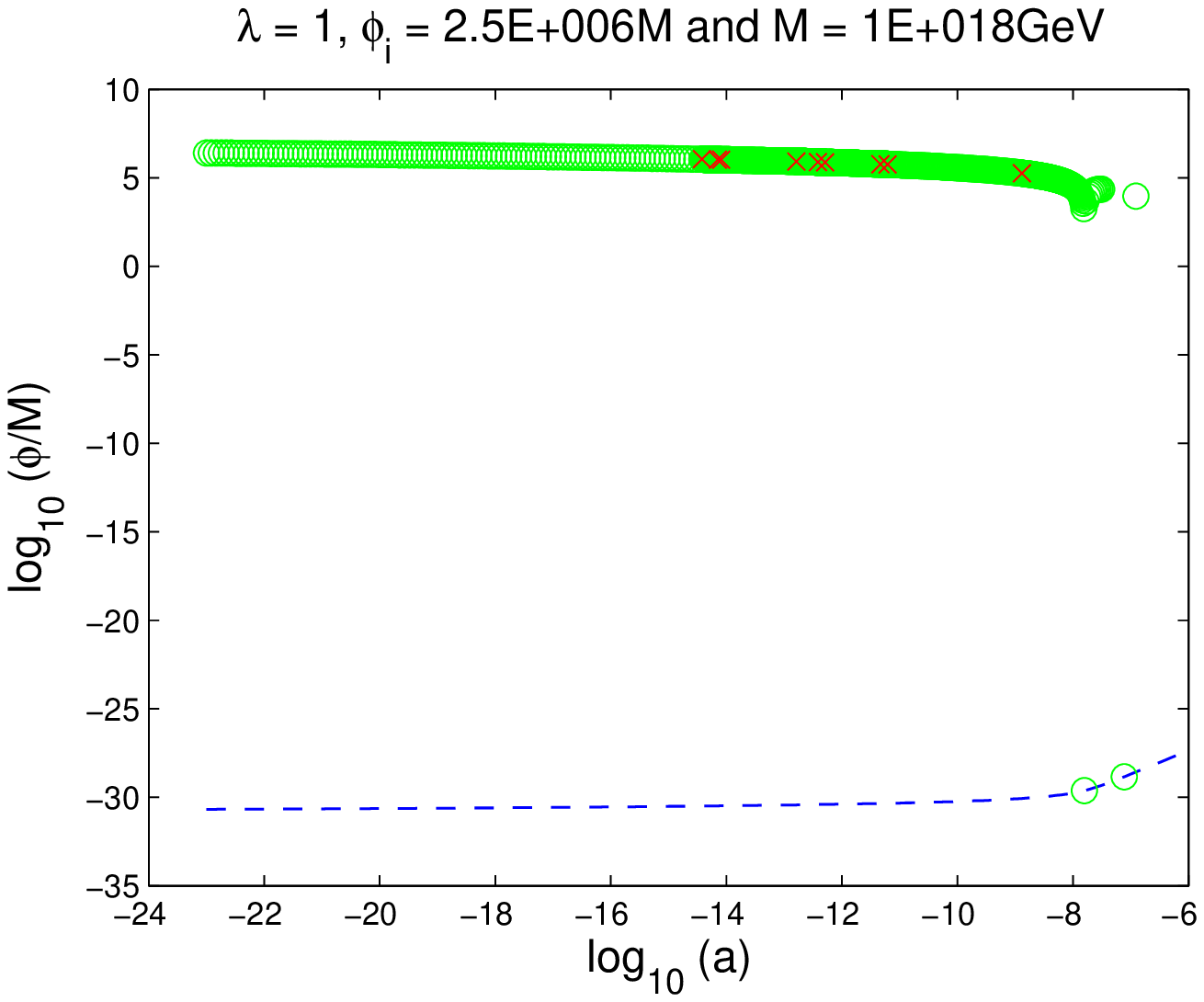}}
\subfigure[$\xi_{B}=10^{-8}$, $\phi_i=10^9M$, $M=1.1\times10^{9}{\rm GeV}$.]{\label{fig:d}\includegraphics[bb=3.5cm 9.1cm 17cm 19.5cm,clip,width=7cm]{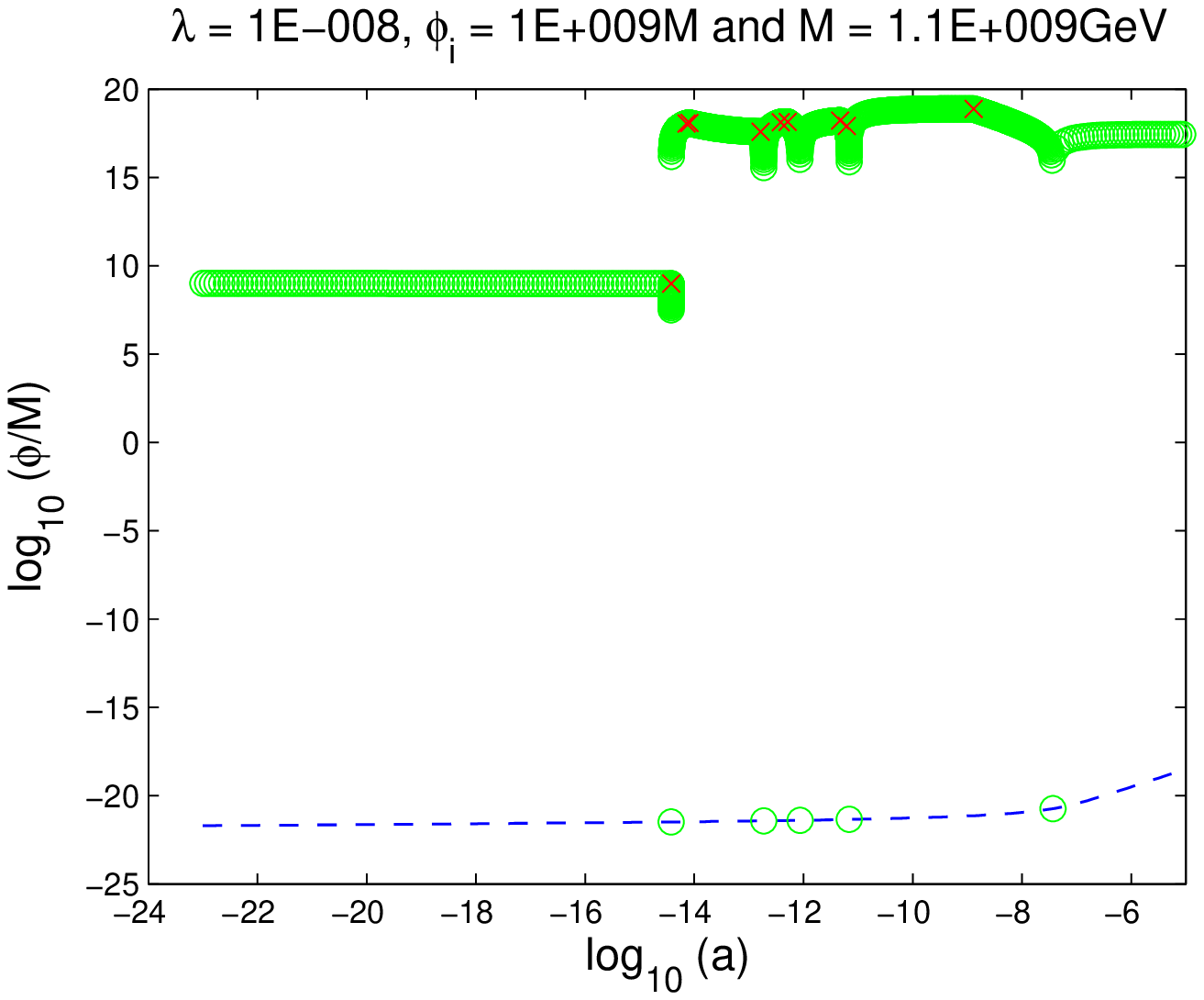}}
\subfigure[$\xi_{B}=10^{-8}$, $\phi_i=1.1M$, $M=10^{18}{\rm GeV}$.]{\label{fig:e}\includegraphics[bb=3.5cm 9.1cm 17cm 19.5cm,clip,width=7cm]{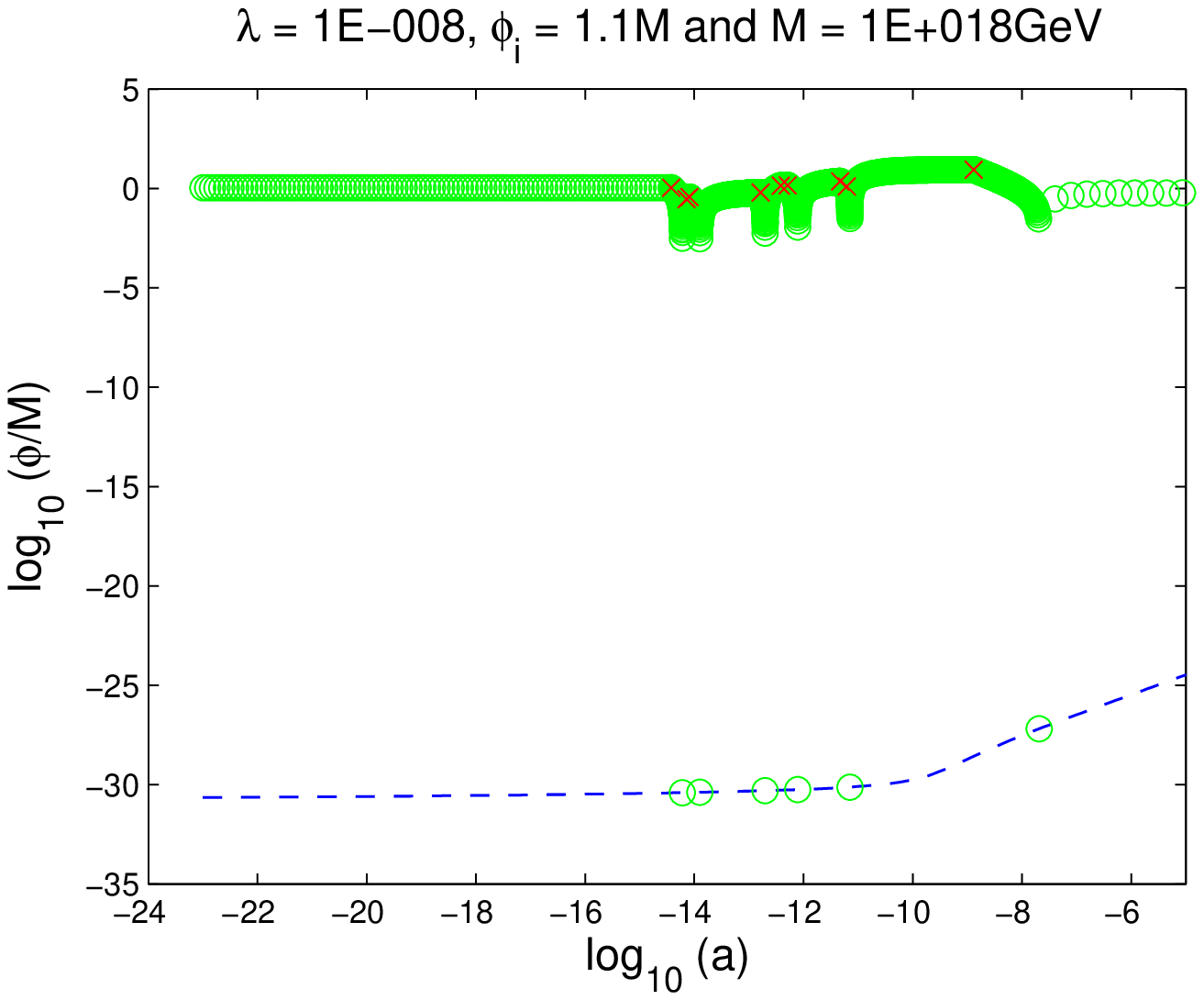}}

\par\end{centering}

\caption{\label{fig:evolution}Chameleon evolution from the end of inflation to BBN, for different $\xi_{B}$, $M$ and $\phi_i$. The dashed blue line is the evolution of $\phi_{\rm min}$. Green circles mark the actual location of the chameleon field including oscillations between its maximum amplitude and the minimum, while the green line is the evolution of $\phi_{\rm max}$ once it enters the `fast oscillations' regime. The red crosses mark the location of the `kicks' as different particle species drop out of thermal equilibrium.}

\end{figure*}

The equations derived in Sections \ref{ApproachingMin} to \ref{sub:kicks} provide a framework for determining the evolution of the chameleon field given a specific starting value at the end of inflation. Figure \ref{fig:evolution} contains a few examples of the resulting prediction for the chameleon evolution after it is released from rest at the end of inflation ($z\sim10^{23}$) through until BBN at $z\sim10^8$, for a variety of parameter values. They were determined through numerical iteration of the methods described in the preceding sections, and we included the contribution from $\rho_{\rm m}$ that was neglected in equation (\ref{eq:EMside}), for the evolution after the electron drops out of thermal equilibrium in the last kick. 

The behaviour of the chameleon is strongly dependent on its starting value and the strength of the chameleon interaction with the primordial magnetic field. The strength of the magnetic interaction depends on both the size of the magnetic field $B_0$ and the coupling strength between the chameleon and electromagnetic field $1/M_{\rm F}$. The bounds on the strength of the primordial magnetic field were discussed in the introduction: observations of the CMB limit the magnitude of $B_0$ to be no greater than approximately $5{\rm nG}$. The strength of the electromagnetic interaction in the chameleon model is constrained by observations of starlight polarization in our galaxy to $M_{\rm F} \gtrsim 1.1\times10^{9}{\rm GeV}$ \citep{Burrage08} (see Section \ref{sec:TheChameleonModel}). The chameleon evolution after inflation depends on the parameter $\beta$, defined in equation (\ref{eq:beta}), and as such is degenerate between $B_0$ and $M_{\rm F}$. We define
\begin{equation*}
\xi_{B}\equiv\left(\frac{B_0}{5{\rm nG}}\right)^2\left(\frac{M_{\rm F}}{1.1\times 10^{9}{\rm GeV}}\right)^{-1}
\end{equation*}
to characterise this dependence. Given the constraints on $B_0$ and $M_{\rm F}$, it is only realistic to consider the range $\xi_{B}\lesssim 1$. We assume the following values for the cosmological parameters: $h\simeq0.7$, $\Omega_{m0}h^2\simeq0.13$ and $\Omega_{r0}h^2\simeq2.5\times10^{-5}$ \citep{Komatsu10}, where $H_0 \equiv 100 h\,{\rm km}\, {\rm sec}^{-1}\,{\rm Mpc}^{-1}$, and we take $n \simeq 2$ in the model for the self-interaction potential.

The plots in Figure \ref{fig:evolution} illustrate the two distinct regimes that the chameleon evolution falls into, depending on the values we choose for the different parameters. For larger magnetic fields and/or when the chameleon lies closer to its minimum at the end of inflation, we find that the chameleon starts to roll towards the minimum of its potential and begins oscillating. See for example Figures \ref{fig:a} and \ref{fig:b}. Often the behaviour satisfies the `fast oscillations' regime so that the amplitude of the chameleon oscillations decay according to the power law defined in equation (\ref{eq:fastOSC}). In this scenario the effect of the kicks is to boost the chameleon into the `fast oscillations' regime, or to have no effect if the field is already oscillating rapidly. 

For smaller magnetic field values, or when the chameleon is released far away from its minimum, we find that the field remains stuck at its initial value until species start dropping out of thermal equilibrium and push the chameleon towards the minimum. This process, however, is highly unstable. If the kick is sufficient to overshoot the minimum then the chameleon effectively bounces off the steep lefthand side of the potential and can shoot back to even larger field values. The chameleon to matter coupling strength described by $M$ determines the energy introduced to the system by the kicks. We saw in equation (\ref{eq:jumpInPhi}) that in the limit $a\gg a_{k}$, $\Delta\phi = - \kappa_k M_{\mathrm{Pl}}\Gamma$, where $\kappa\propto 1/M$. Thus the stronger the matter coupling, the greater will be the jump in the $\phi$ evolution as a result of the kicks. For example, Figures \ref{fig:d} and \ref{fig:e} have the same $\xi_{B}$ and $\phi_i$ but two choices of $M$. The smaller $M$ value sends the field to much larger values as a result of the kicks. It is this second regime that exists in the case of no primordial magnetic field or when the electromagnetic interaction of the chameleon is ignored. For this reason, the initial conditions considered by Brax et al. \cite{Brax05} had to be finely tuned so that the chameleon ended up near the minimum of its potential at the onset of BBN. 

The ratio of $\phi_i$ to $\xi_{B}$ dictates whether the chameleon has started to oscillate before the first kick when the top quark drops out of thermal equilibrium. We approximate the transition between these two regimes by the requirement that the field has oscillated exactly once between its release at the end of inflation and the first kick. The oscillation period is given by equations (\ref{eq:amin}) and (\ref{eq:a2}). In the limit $a_{2}\gg a_1$ we can approximate this by
\begin{equation*}
\frac{a_2}{a_1}\simeq 2\exp\left(\frac{\phi_1}{\beta}+1\right).
\end{equation*}
Substituting for $a_1 = a_i\sim10^{-23}$ and $a_2=a_{\rm kick}\sim3.77\times10^{-15}$, leads to the condition $\phi_1\approx18\beta$, which implies the transition occurs for
\begin{equation}
\phi^{\rm (crit)}_i \approx 3.8\,\xi_{B}\times 10^5 M_{\rm Pl}.\label{eq:Lbound} 
\end{equation}
When the chameleon field value at the end of inflation is less than $\phi^{\rm (crit)}_i$ the chameleon will immediately begin to roll towards the minimum of its potential and start oscillating. For field values greater than $\phi^{\rm (crit)}_i$ the chameleon is effectively frozen at this position until particle species begin dropping out of thermal equilibrium. In the next section we discuss the implication of these results on the ability of the chameleon to satisfy constraints on the variation of particle masses after BBN.

\section{BBN Constraints}\label{BBNconstraints}

The analysis of Brax et al. \cite{Brax05}, which neglected the chameleon interaction to electromagnetism, imposed a constraint of 10\% on the allowed variation of particle masses after BBN. This places a limit on the maximum amplitude of the chameleon oscillations at the onset of BBN. The  bound on the allowed variation of particle masses coming from big bang nucleosynthesis is sensitive to a number of parameters which are not well understood, and so the constraint of 10\% does not reflect a hard experimental bound. We discuss the BBN constraints in more detail below, but it is beyond the scope of this paper to perform a full analysis. Instead, we investigate the effects on the chameleon model of imposing a fractional constraint, $\gamma_{\rm BBN}$, on the allowed variation of particle masses, and explicitly consider the case $\gamma_{\rm BBN} = 0.1$ for a direct comparison to the results of Brax et al. \cite{Brax05}. 
  
The BBN constraints are relevant because the amplitude of the chameleon oscillations about its minimum can potentially result in a large variation in the masses of fundamental particles. The matter particles follow geodesics of the conformal metric, $A_{\rm m}^2\left(\phi\right)g_{\mu\nu}$. In the conformal (Jordan) frame, in which the chameleon is non-minimally coupled to gravity, the particle masses $m_J$ are independent of $\phi$. However in the physical Einstein frame they acquire a $\phi$ dependence: $m(\phi)=A_{\rm m}(\phi)m_{J}$ \citep{FujiiMaeda}. Variation in $\phi$, therefore, leads to a variation in the observed mass of particles:
\begin{equation*}
\left|\frac{\Delta m(\phi)}{m(\phi)}\right| = \frac{A_{\rm m}^{\prime}(\phi)}{A_{\rm m}(\phi)}\left|\Delta\phi\right| = \frac{1}{M}\left|\Delta\phi\right|.  
\end{equation*}  
The earliest time at which cosmological observations can constrain the variation in particle masses is at big bang nucleosynthesis. The primordial element abundances are sensitive to variation in the parameters of the Standard Model of particle physics. Dent et al. \cite{Dent07,Dent08} investigated the response of the element abundances to individual variations in the fundamental parameters as well as considering unified models which combined the variation of more than one parameter. Varying only one of the parameters at a time, they were able to place $2\sigma$ bounds on the allowed variation for the different fundamental couplings from BBN until today. The allowed range for the mass of the electron was found to be $-17\%\leq \Delta\ln m_{\rm e}\leq +9\%$. Although this result neglects the complex interactions from multiple fundamental parameters varying, it allowed Brax et al. \cite{Brax05} to estimate a constraint on the variation in particle masses arising from the chameleon model to be less than approximately 10\%. In our analysis we generalise this constraint to some fraction $\gamma_{\rm BBN}$, which imposes 
\begin{equation}
\phi^{\rm (BBN)} \lesssim \gamma_{\rm BBN} M,\label{eq:BBNbound}
\end{equation}
since $\phi^{\rm (today)}_{\rm min}\approx 10^8\Lambda \ll \gamma_{\rm BBN} M$. The strength of the matter coupling described by $M$ has a significant impact on whether the constraints at BBN are satisfied. There are subtler issues of whether the chameleon itself influences the primordial element abundances; see \cite{Steigman07} for a general review of primordial nucleosynthesis and the effects of non-standard physics. However, as first shown by \cite{Brax05}, the deviations of the chameleon model from standard LCDM are insignifcant at the scales relevant for the expansion rate and baryon/photon ratio.

\begin{figure}
\begin{centering}
\includegraphics[width=7.5cm]{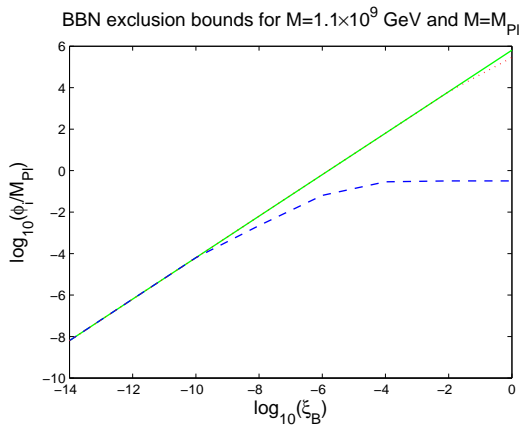}
\par\end{centering}

\caption[The parameter space excluded by BBN constraints.]{\label{fig:BBNbounds}The parameter space of $\phi_i$ and $\xi_{B}$ that is excluded by BBN constraints when $\gamma_{\rm BBN}=0.1$. The dashed blue line marks the excluded region for $M=1.1\times10^9{\rm GeV}$, and the dotted red for $M=M_{\rm Pl}$. The allowed parameter space lies to the bottom right of the exclusion lines. The green line marks the transition from the chameleon oscillating prior to the `kicks', and the one in which the field is frozen prior to the `kicks'. It is independent of $M$.}

\end{figure}

In the previous section we were able to categorise the chameleon behaviour into two distinct groups depending on whether the parameter values were such that the chameleon was oscillating about the minimum of its potential prior to the top quark dropping out of thermal equilibrium, or whether the field was frozen at its initial value. It is this second scenario that occurs when there is no interaction between the chameleon and electromagnetic field, and corresponds to the situation analysed by Brax et al. \cite{Brax05}. When the particle species drop out of thermal equilibrium they introduce a a short-lived boost of energy into the system and `kick' the chameleon field towards its minimum. The initial field value of the chameleon needs to be strongly fine-tuned for the cumulative effect of the kicks to result in the field being positioned close to the minimum of its potential, rather than overshooting and rolling back to even larger field values, as shown by the plots in Figures \ref{fig:c} to \ref{fig:e}. We find than in general, whenever the chameleon starts out frozen at its initial value, it is impossible to satisfy the constraints on the chameleon field at BBN for all $\gamma_{\rm BBN} < \mathcal{O}(1)$ without extreme fine-tuning. Only if the allowed variation of particles is a few 100\% will the chameleon fall below the appropriate threshold at BBN. The parameter space boundary that separates the two regimes was derived in equation (\ref{eq:Lbound}). For a given initial field value $\phi_i$, this implies the following constraint on $B_0$ and $M_{\rm F}$ for all $\gamma_{\rm BBN} < \mathcal{O}(1)$:
\begin{eqnarray}
\left(\frac{B_0}{5{\rm nG}}\right)^2\left(\frac{M_{\rm F}}{1.1\times 10^{9}{\rm GeV}}\right)^{-1}&\gtrsim& 2.4\times10^{-6}\frac{\phi_i}{M_{\rm Pl}}. \label{eq:constraint1}
\end{eqnarray}
Even when the primordial magnetic field large enough to cause the chameleon to oscillate prior to the onset of the kicks, there is a limit on the speed with which the amplitude of the oscillations can decay, which further constrains the parameters. We consider the case when the chameleon is in the fast oscillations regime from the very start of its evolution. In this situation the chameleon converges most rapidly on the minimum following the power-law behaviour described by equation (\ref{eq:fastOSC}). Although it requires a large ratio of $\beta/\phi_i$ to place us in this regime, the convergence is independent of $\beta$. If the initial conditions for the field satisfy $\phi_i/\beta\ll1$, then the subsequent evolution is determined by
\begin{eqnarray*}
\log\left(\frac{\phi_{\rm max}}{\phi_i}\right) & = & -\frac{2}{3}\log\left(\frac{a}{a_{i}}\right)
\end{eqnarray*} 
and so $\phi_{\rm max}^{\rm (BBN)} \approx 10^{-10}\phi_i$.
To satisfy the BBN constraints, this imposes
\begin{equation}
\phi_i\lesssim 10^{10}\gamma_{\rm BBN} M. \label{eq:Mbound}
\end{equation}

In Figure \ref{fig:BBNbounds} are the exclusion bounds determined empirically from requiring the chameleon to fall below the $\gamma_{\rm BBN} = 0.1$ threshold at BBN. These are presented for two different choices of the matter coupling strength: $M=1.1\times10^9{\rm GeV}$ and $M=M_{\rm Pl}$, motivated by the constrained value of $M_{\rm F}$ assuming $M\sim M_{\rm F}$ and a natural gravitational scale respectively. The bounds agree closely with the limiting cases given in equations (\ref{eq:constraint1}) and (\ref{eq:Mbound}). The transition between when the chameleon is oscillating and when it is frozen at its initial value marks the BBN exclusion bound for small $\phi_i$ and, as we can see from Figure \ref{fig:BBNbounds}, there is a saturation point at $\phi_i \sim 1M_{\rm Pl}$ for the case of $M=1.1\times10^9\,{\rm GeV}$, which is dependent on $M$ but independent of $\xi_{B}$, which excludes larger values of $\phi_i$. There is a small range of intermediate $\xi_{B}$ values for which the field oscillates more than once before the first kick, but is not actually in the fast oscillations regime from the start of its evolution (Figure \ref{fig:b} for example), which forms a smooth transition between the two limiting cases in the exclusion plot.

The bounds in equations (\ref{eq:constraint1}) and (\ref{eq:Mbound}) lead to a constraint on $B_0$, $M_{\rm F}$ and $M$ if we impose a range of $\phi_{i}$ that we consider likely for the chameleon field at the end of inflation. They only apply if the constraints at BBN limit the variation of particle masses after BBN to be less than $\mathcal{O}(1)$, but they are not sensitive to the exact value of 10\% imposed by Brax et al. \cite{Brax05}. We consider a few different physical energy scales which could act as natural initial conditions for the chameleon field at the end of inflation. The resulting bounds are,
\begin{eqnarray*}
\phi_i\sim&\mathcal{O}(M_{\rm Pl}), & \textstyle \left(\frac{B_0}{5{\rm nG}}\right)^2\left(\frac{M_{\rm F}}{1.1\times 10^{9}{\rm GeV}}\right)^{-1} \gtrsim 2.4\times10^{-6}, \\
\phi_i\sim&\mathcal{O}(\Lambda_{\rm EW}), &  \textstyle\left(\frac{B_0}{5{\rm nG}}\right)^2\left(\frac{M_{\rm F}}{1.1\times 10^{9}{\rm GeV}}\right)^{-1}\gtrsim 6.4\times10^{-22}, \\
\phi_i\sim&\mathcal{O}(\Lambda_{\rm QCD}), &  \textstyle\left(\frac{B_0}{5{\rm nG}}\right)^2\left(\frac{M_{\rm F}}{1.1\times 10^{9}{\rm GeV}}\right)^{-1}\gtrsim 5.3\times10^{-25}, \\
\phi_i\sim&\mathcal{O}(\Lambda_{\rm SUSY}), &  \textstyle\left(\frac{B_0}{5{\rm nG}}\right)^2\left(\frac{M_{\rm F}}{1.1\times 10^{9}{\rm GeV}}\right)^{-1}\gtrsim 2.6\times10^{-22},
\end{eqnarray*} 
where the electroweak energy scale $\Lambda_{\rm EW}\sim 246{\rm GeV}$, the QCD energy scale $\Lambda_{\rm QCD}\sim 200{\rm MeV}$ and we assume some supersymmetric breaking scale with $\Lambda_{\rm SUSY}\sim 1{\rm TeV}$. The constraints on $M$ with $\gamma_{\rm BBN}\lesssim \mathcal{O}(1)$, are already trivially satisfied by experiment for all but $\Oo(M_{\rm Pl})$ initial conditions. When $\phi_i\sim\Oo(M_{\rm Pl})$ this imposes $M \gtrsim 10^{8}\gamma_{\rm BBN}^{-1}{\rm GeV}$.

We can compare these constraints to the bounds placed on $B$ and $M_{\rm F}$ from considering CMB photons mixing with chameleons in the presence of a primordial magnetic field. In \cite{Schelpe10} the effects on the CMB of photon-scalar mixing in the PMF along the path from recombination to reionization were analysed. The predicted modification to the CMB intensity spectrum was compared to high precision measurements of the CMB monopole made by the FIRAS instrument on board the COBE satellite. This implied a degenerate bound in the range $(B_{0}/1{\rm nG})\left(M_{\rm F}/10^9{\rm GeV}\right)^{-2}\lesssim
3\times10^{-6}$ to $(B_{0}/1{\rm nG})^{1/2}\left(M_{\rm F}/10^9{\rm GeV}\right)^{-2}\lesssim
2\times10^{-8}$, at 95\% confidence. The range depends on the PMF spectral index, assumed to run from $n_{\rm B}=-2.9$ to $-1.0$. These bounds can be combined with the above BBN constraints on $B_0$ and $M_{\rm F}$. Assuming order $M_{\rm Pl}$ initial conditions for the chameleon in the early Universe, and taking $n_B=-2.9$ (which gives the most conservative bounds on the magnetic field strength), we find the following allowed band for the chameleon-photon coupling strength, 
\begin{equation*}
600\left(\frac{B_{0}}{1\mathrm{nG}}\right)^{\frac{1}{2}}\lesssim\left(\frac{M_{\mathrm{F}}}{10^{9}\mathrm{GeV}}\right)\lesssim 1.6\times10^{4}\left(\frac{B_{0}}{1\mathrm{nG}}\right)^{2}.
\end{equation*}
For primordial magnetic fields smaller than $10^{-10}{\rm G}$, there are no coupling strengths that can satisfy these combined bounds. We note that the allowed degree of chameleon-photon mixing in the early Universe places an upper bound on the PMF strength and chameleon-electromagnetic interaction, while for the chameleon to be well behaved in the early Universe and satisfy constraints at BBN requires a much stronger magnetic interaction to drive the chameleon to its minimum in time. There is only a narrow band of parameter values that satisfy these constraints and in all cases they require $B_0 \gtrsim 0.1{\rm nG}$. For different values of the magnetic field spectral index the bounds become even stronger with $B_0 \gtrsim 2{\rm nG}$ for $n_B = -1.0$. 

These bounds hold for any $<\mathcal{O}(1)$ constraint on the variation of particle masses after BBN. They imply that a sizable primordial magnetic field ($\sim 0.1{\rm nG}$) must exist if the chameleon is to satisfy these constraints at BBN, and if it is released from an arbitrary position within its potential at the end of inflation that lies within some natural range ($\phi_{i}\lesssim M_{\rm Pl}$). Without a primordial magnetic field we are reduced to the solution considered by Brax et al. \cite{Brax05} which required extreme fine-tuning of the initial conditions. In this situation the chameleon field is frozen at its initial value, and is only pushed towards the minimum by particle species dropping out of thermal equilibrium. The rebound at the minimum of the potential can cause the chameleon to shoot to even larger field values by the end of the kicks, and so fine-tuning is needed to satisfy the BBN constraints.


\section{Conclusions}\label{EarlyUniverseDiscussion}

When the chameleon model was first suggested by Khoury and Weltman \cite{Khoury04,Khoury03}, the authors in \cite{Brax05} examined the cosmological evolution of the chameleon and determined that, for certain specific starting values for the field around $\Oo (M_{\rm Pl})$ at the end of inflation, constraints at BBN can be satisfied. However the extreme fine-tuning necessary for these initial conditions is unsatisfactory if the chameleon is to provide a natural candidate for dark energy. In \cite{Brax10} it was shown that the coupling of the scalar field to electromagnetism should not be neglected in these models. Not only does the electromagnetic interaction of the chameleon lead to the prediction of new phenomena such as chameleon-photon mixing in background magnetic fields, it also allows the possibility of a primordial magnetic field modifying the early Universe behaviour of the chameleon.  

In this article, we have demonstrated that the presence of a primordial magnetic field can help drive the chameleon field to the minimum of its potential. Without a primordial magnetic field (or at least with a very weak one), the chameleon is frozen at its initial value at the end of inflation and requires the energy from particle species dropping out of thermal equilibrium to drive it towards the minimum. However, this process is very unstable since the energy introduced by particle species going non-relativistic is akin to a delta function and easily causes the field to overshoot the minimum and oscillate more violently. By contrast, a strong PMF causes the field to oscillate about the minimum of its potential with a decaying amplitude as the Universe expands, and can satisfy constraints at BBN for a range of initial conditions.

The paper by Brax et al. \cite{Brax05} that originally examined the cosmological evolution of the chameleon in the case of zero EM coupling, imposed an upper limit on the chameleon field at BBN arising from a 10\% constraint on the variation of particle masses. This constraint was suggested by Dent et al. \cite{Dent07,Dent08} based on an anaylsis of the primordial element abundances at BBN. However, the degeneracies between the different factors influencing the element abundances are not well understood and so this can only be taken as an estimate. 

In this work we have developed a semi-analytical solution to the early Universe evolution of the chameleon as it falls towards the minimum of its potential, allowing for the coupling of the chameleon to any electromagnetic fields. This provides a valuable framework for anyone wishing to analyse the evolution of the chameleon in the light of new experimental bounds that may be placed upon the theory. Although the 10\% constraint on particle masses at BBN is only an estimate, and a full analysis of the BBN constraints is beyond the scope of this paper, we tentatively suggest that a BBN constraint on the allowed variation of particle masses being less than 100\% is not unreasonable. If we assume the natural scale for inflationary scenarios is around the Planck mass, then a natural value for the chameleon at the end of inflation is up to and including $\phi_i\sim\Oo(M_{\rm Pl})$. Combined with the tentative $<100\%$ constraint at BBN, this places a degenerate constraint on the PMF strength and chameleon-photon coupling, 
$$\left(\frac{B_0}{5{\rm nG}}\right)^2\left(\frac{M_{\rm F}}{1.1\times 10^{9}{\rm GeV}}\right)^{-1} \gtrsim 2.4\times10^{-6}.$$ The degeneracy of the constraint can be broken by considering the effects of chameleon-photon mixing in the primordial magnetic field \citep{Schelpe10}. This imposes a narrow band of allowed parameters which can only be satisfied for $B_0\gtrsim 0.1 {\rm nG}$. These magnetic field values are close to the level that could be detected or ruled out by the next generation of CMB experiments. We also find that the chameleon to matter coupling strength is constrained by the limit at BBN to have 
$$M\gtrsim 10^8{\rm GeV},$$ 
which is many orders of magnitude greater than the existing direct constraints on $M$. 

Our analysis has been with specific reference to the chameleon model \cite{Khoury04,Khoury03}. However many theories of modified gravity are required to have some form of chameleon mechanism if they are to satisfy the many weak-field limit tests and explain dark energy or dark matter \cite{Jain10}. It would be interesting to investigate the role that can be played by primordial magnetic fields in these alternative theories, in the light of our results.

\section*{Acknowledgments} We thank Douglas Shaw for all his help and advice during the early stages of this project, and we are grateful to Carsten van de Bruck, Clare Burrage, Anne Davis and Hans Winther for useful discussions. DFM thanks the funding from the Research Council of Norway. DFM is also partially supported by project PTDC/FIS/111725/2009 and CERN/FP/116398/2010.

\end{document}